\documentclass[twocolumn,aps,prd,amsmath,amssymb,preprintnumbers,longbibliography]{revtex4-1}
\usepackage{amsmath} \usepackage{amsfonts} \usepackage{amssymb}
\usepackage{bbm}
\usepackage{epsfig}
\usepackage{graphics}
\usepackage{graphicx}
\usepackage{titlesec}
\usepackage{mathtools}
\usepackage{environ}
\usepackage{dsfont}
\usepackage{version}
\usepackage[colorlinks=true]{hyperref}
\hypersetup{urlcolor=black,linkcolor=black,citecolor=black}
\usepackage{natbib}
\usepackage[toc,page]{appendix}

\usepackage{enumerate}

\usepackage{dsfont}

\makeatletter
\renewcommand*\env@matrix[1][c]{\hskip -\arraycolsep
  \let\@ifnextchar\new@ifnextchar
  \array{*\c@MaxMatrixCols #1}}
\makeatother

\makeatletter										
\def\p@subsection{\thesection .\,} 
\makeatother                                       

\newcommand{\be}{\begin{equation}}
\newcommand{\ee}{\end{equation}}
\newcommand{\ba}{\begin{align}}
\newcommand{\ea}{\end{align}}
\newcommand{\nn}{\nonumber}

\newcommand{\gl}{\big(}
\newcommand{\gr}{\big)}

\newcommand{\vp}{{\varphi}}
\newcommand{\K}{{\mathcal K}}

\newcommand{\te}{t+\varepsilon}

\titleformat{\subsection}[block]{\normalfont\bfseries}{\thesubsection.}{1ex}{}
\titlespacing{\subsection}{0pt}{10pt}{1pt}[0pt]
\titleformat*{\section}{\large\bfseries}
\renewcommand{\thesubsection}{\arabic{subsection}}

\pdfoutput=1 

\usepackage{todonotes}
\usepackage{braket}

\newcommand{\cL}{\mathcal{L}}
\newcommand{\cM}{\mathcal{M}}
\newcommand{\psihat}{\widehat{\psi}}

\newcommand{\bel}[1]{\be\label{#1}}

\newcommand{\Ktil}{\tilde{\K}}

\newcommand{\subt}[1]{_{\text{#1}}}

\newcommand{\inn}{\subt{in}}

\newcommand{\qq}[1]{``#1"}

\newcommand{\Shat}{\widehat{S}}

\newcommand{\eps}{\varepsilon}

\newcommand{\pmat}[4]{\begin{pmatrix} #1 & #2 \\ #3 & #4 \end{pmatrix}}
\newcommand{\pvec}[2]{\begin{pmatrix} #1 \\ #2 \end{pmatrix}}

\newcommand{\g}{_{\gamma}}

\newcommand{\mo}{^{\mu}}

\newcommand{\mb}{_{\mu}}

\newcommand{\chiu}[2]{\chi_{#1}^{#2}}
\newcommand{\chipu}[2]{{\chi'}_{#1}^{#2}}

\newcommand{\Ttil}{\widetilde{T}}
\newcommand{\chibar}{\bar\chi}
\newcommand{\Itil}{\tilde{I}}

\newcommand{\Lbar}{\bar L}

\definecolor{refkey}{rgb}{0,0,1}
\definecolor{labelkey}{rgb}{0,1,0}

\begin{document}


\title{\LARGE Cellular automaton for spinor gravity in four dimensions}

\author{C. Wetterich}

\affiliation{Institut  f\"ur Theoretische Physik\\
Universit\"at Heidelberg\\
Philosophenweg 16, D-69120 Heidelberg}

\begin{abstract}
Certain fermionic quantum field theories are equivalent to probabilistic cellular automata, with fermionic occupation numbers associated to bits. We construct an automaton that represents a discrete model of spinor gravity in four dimensions. Local Lorentz symmetry is exact on the discrete level and diffeomorphism symmetry emerges in the naive continuum limit. Our setting could serve as a model for quantum gravity if diffeomorphism symmetry is realized in the true continuum limit and suitable collective fields for vierbein and metric acquire non-vanishing expectation values. The discussion of this interesting specific model reveals may key qualitative features of the continuum limit for probabilistic cellular automata. This limit obtains for a very large number of cells if the probabilistic information is sufficiently smooth. It is associated to coarse graining. The automaton property that every bit configuration is updated at every discrete time step to precisely one new bit configuration does no longer hold on the coarse grained level. A coarse grained configuration of occupation numbers can evolve into many different configurations with certain probabilities.
This characteristic feature of quantum field theories can come along with the emergence of continuous space-time symmetries.
\end{abstract}

\maketitle

\section{Introduction}

Probabilistic automata are quantum systems in a discrete real formulation for the time evolution. At every given time $t$ we consider bit configurations $\rho$. The probabilistic information is encoded in a real wave function $q_{\rho}(t)$, with $p_\rho(t)=q_\rho^2(t)$ the probability to find at $t$ the bit configuration $\rho$. The updating to a new bit configuration $\tau$ at the next time step $t+\eps$ is given by the evolution law
\bel{INN1}
q_\tau(t+\eps)=\gl\Shat(t)\gr_{\tau\rho}q_\rho(t)\ .
\ee
The step evolution operator $\Shat(t)$ is a unique jump matrix with a unique element one in each row and column, and zero elements otherwise,
\bel{INN2}
\Shat_{\tau\rho}=\delta_{\tau,\bar\tau(\rho)}=\delta_{\bar\rho(\tau),\rho}\ .
\ee
The map $\rho\to\bar\tau(\rho)$ defines the updating map from $\rho$ to $\bar\tau(\rho)$, where we assume invertibility with $\bar\rho(\tau)$ the inverse map. The probability distribution is updated with the same step evolution operator, such that the probability $p_\tau(t+\eps)$ coincides with the probability $p_{\bar\rho(\tau)}(t)$ for the configuration $\bar\rho(\tau)$ from which it originates at the previous time $t$. One can change the sign of elements of $\Shat$ without affecting the updating of the probabilities.

The probabilistic automata discussed here have a deterministic time evolution, with a probabilistic nature only due to probabilistic initial conditions at some initial time $t\inn$. One can define an \qq{overall probability distribution} for bit configurations at all times as a type of generalized Ising model~\cite{CWFGI, CWPW}, for which the probabilities vanish for all time-neighboring configurations not allowed by the updating rule. The probabilistic information is encoded in boundary terms involving only the bit configurations at initial and final time. The overall probabilities are positive for arbitrary signs of the components $q_\tau(t)$. The overall probability distribution defines a classical statistical system. One only employs the standard laws for \qq{classical} probabilities of the overall system. The \qq{quantum axioms} follow from a projection on the \qq{time-local subsystem}~\cite{CWPW}.

Two simple steps have to be taken in order to realize standard continuous complex quantum mechanics. The first concerns a complex structure. We assume the presence of an involution $K$ such that the $N$-component vector $q(t)$ can be divided in an $N/2$-component vector $q_R(t)$ which is even under $K$, and another $N/2$-component vector $q_I(t)$ which is odd under $K$. The equal size of the even and odd parts is guaranteed by a second map $I$, $I^2=-1$, $\{K,I\}=0$. This pair of discrete transformations allows for a map from the $N$-component real wave function $q$ to the $N/2$-component complex wave function $\vp$,
\bel{INN3}
q\to\vp(q)\ ,\quad \vp(t)=q_R(t)+iq_I(t)\ ,
\ee
with the properties
\bel{INN4}
\vp(Kq)=\vp^*(q)\ ,\quad \vp(Iq)=i\vp(q)\ .
\ee

A rather general involution $K$ suitable for our purpose is particle-hole conjugation~\cite{FPCA}. This interchanges the values one and zero of the bits, or maps corresponding to occupation numbers $n\g(t)\to\gl1-n\g(t)\gr$, or flips the sign of Ising spins $s\g(t)=2n\g(t)-1$. We focus on particle-hole invariant updatings for which $\Shat$ is block diagonal, acting in the same way on $q_R$ and $q_I$. This results in the discrete complex evolution equation
\bel{IN5}
\vp(t+\eps)=\Shat_R(t)\vp(t)\ ,
\ee
with real $N/2\times N/2$ matrix $\Shat_R$. A more general step evolution operator $\Shat=\Shat_R\otimes\mathds{1}_2+\Shat_II$ remains compatible with the complex structure as well, defining the complex matrix
\bel{INN6}
U(t)=\Shat_R(t)+i\Shat_I(t)\ .
\ee

The second step concerns a formulation in continuous time. Unique jump matrices are orthogonal matrices,
\bel{INN6A}
\Shat(t)\Shat^T(t)=1\ .
\ee
In the complex formulation the corresponding $N/2\times N/2$-matrices~\eqref{INN6} are unitary matrices. Combining a certain number of time steps from $t$ to $t+\Delta t$ results in the unitary evolution operator
\begin{align}
\label{INN7}
U(t+\Delta t,t)=&\,U(t+\Delta t-\eps)\dots U(t+\eps)U(t)\ ,\nn\\
\vp(t+\Delta t)=&\,U(t+\Delta t,t)\vp(t)\ .
\end{align}
This evolution operator can be used to define a hermitian Hamiltonian
\bel{INN8}
U(t+\Delta t)=\exp\Big\{-i\Delta tH(t)\Big\}\ ,\quad H^\dagger(t)=H(t)\ .
\ee
As a consequence, the time evolution obeys the continuous Schrödinger equation
\bel{INN9}
i\partial_t\vp(t)=H(t)\vp(t)\ ,
\ee
where $H(t)$ is taken piecewise constant in the interval between $t$ and $\Delta t$. Indeed, the solution of eq.~\eqref{INN9} reproduces eq.~\eqref{INN7} for discrete time points $t=t\inn+n\Delta t$ with integer $n$. It can be taken as a suitable interpolation for continuous time. This closes the formal argument that probabilistic automata evolve as quantum systems. We will be interested in situations where characteristic time scales are much longer than $\Delta t$.

We consider at every $t$ configurations of a large number of bits $n\g(t)=(1,0)$, where $\gamma$ denotes both discrete positions in space $\vec x$ and \qq{internal quantum numbers} $\alpha$, $\gamma=(\vec x,\alpha)$, such that $n\g(t)=n_\alpha(t,\vec x)$. We identify the position $\vec x$ with the cells of a cellular automaton. The \qq{updating} from $t$ to $t+\eps$ of the bit configuration $\{n_\alpha(t,\vec x)\}$ within a given cell $\vec x$ is influenced only by the bit configurations of a few neighboring cells at $t$. Analogously, the bit configuration within a cell $\vec x$ at $t+\eps$ influences only the state of a few neighboring cells at $t+2\eps$. This cellular property generates backward and forward \qq{light cones} and therefore the characteristic causal structure of local quantum field theories. Probabilistic cellular automata of this type can be interpreted as local quantum field theories.

More generally, cellular automata encode a ``unitary time evolution" where no information is lost in a discrete setting with locality properties~\cite{JVN,ULA,ZUS,HPP,TOOM,DKT,VICH,FLN,WOLF,CREU,TOMA,LIRO,HED,GAR,PREDU,RICH,AMPA}. They are the basis for interesting attempts to obtain quantum mechanics from deterministic physics~\cite{GTH,ELZE,HOOFT2,HOOFT3,HOOFT4}. For a deterministic automaton a given bit configuration $\bar\rho$ at $t$ corresponds to a sharp wave function with elements $q_\rho(t)=\pm\delta_{\rho\bar\rho}$. This is mapped at $\te$ to a unique configuration $\bar\tau(\bar\rho)$ by $q_\tau(\te)=S_{\tau\rho}(t)q_\rho=S_{\tau\bar\rho}(t)=\pm\delta_{\tau\bar\tau}$, where $\bar\tau$ corresponds to the unique non-zero element $S_{\bar\tau\bar\rho}(t)$ associated to $\bar\rho$. A probabilistic automaton is defined by a probability distribution for the initial bit configurations, with the probability $p_\rho(t)$ to find at $t$ the bit-configuration $\rho$. The relation $p_\rho(t)=q_\rho^2(t)$ introduces the concept of a wave function for classical probabilistic systems~\cite{CWQPCS}.

We are interested in the continuum limit for which characteristic time and length scales are much larger than $\Delta t$. This assumes a sufficiently smooth wave function $\vp(t)$ both in time and space. For our cellular automata this wave function is a multi-particle wave function and the smoothness in space should apply to all particle positions. One-particle wave functions $\vp_\sigma(t,\vec x)$ are only a particular case. Computing the continuum limit for a very large number of cells is a challenging enterprise. In this note we only take first steps by discussing the \qq{naive continuum limit} which will be explained later. In this context the possibility to group a sequence of step evolution operators to a combined evolution operator~\eqref{INN7} is of great help in order to realize continuous rotation symmetry in the naive continuum limit. This can be extended to other continuous symmetries as diffeomorphism invariance. In this note we will discuss a specific model of spinor gravity. The techniques developed here should be viewed as building blocks for the construction of rather general probabilistic cellular automata with an interesting naive continuum limit.

Configurations of bits or Ising spins are equivalent to the states of a multi-fermion system in the occupation number basis. The fermionic occupation numbers $n_\alpha(\vec x)=(1,0)$ for a fermion of type $\alpha$ situated at the position $\vec x$ define a basis for the quantum states of the multi-fermion system. In the occupation number basis the general wave function at time $t$ has components $\vp_\tau(t)$, with $\tau=\{n_\alpha(\vec x)\}$ denoting the possible configurations of occupation numbers. This is precisely the wave function of a probabilistic cellular automaton. For a large total number $M$ of occupation numbers $\gamma=(\vec x,\alpha)$ the number of configurations $\tau$ is huge, $N=2^M$. This underlines that only a probabilistic description is meaningful. As long as $M$ remains finite the space of functions $\vp(t)$ is a finite-dimensional complex vector space restricted to normalized vectors $\sum_\tau\vp_t^*(t)\vp_\tau(t)=1$. A continuum limit for $M\to\infty$ is possible for smooth enough wave functions belonging to a Hilbert space. Probabilistic cellular automata are fermionic quantum field theories.

In the other direction it has been established that some particular discrete fermionic quantum field theories are equivalent to probabilistic cellular automata~\cite{CWPCA,CWNEW,FPCA}. This holds in the sense that all observables have the same expectation value, including all correlations, in both pictures. The evolution of appropriate wave functions is identical as well. The basic idea is simple. The equivalence is demonstrated by computing the step evolution operator for a suitable discretized fermionic quantum field theory. It is sufficient to show that in a real discrete formulation the step evolution operator $\widehat{S}(t)$ is a unique jump matrix. This means that for each discrete time step every configuration of occupation numbers evolves to a single definite new configuration.

For computing the automaton associated to a fermionic quantum field theory we reverse the construction above. Consider a fermion system with Hamiltonian $H$ in the occupation number basis. For discrete time steps from $t$ to $\te$ the unitary evolution operator is given by $U(t)=\exp[-i\eps H(t)]$, where we assume that $H$ is constant in the interval $[t,\te]$. The evolution of the complex wave function $\vp(t)$ can always be written in a real formulation as
\begin{align}
\label{I1}
\vp(\te)=&U(t)\vp(t)\ ,\quad q(\te)=\Shat(t)q(t)\ ,\nn\\
q=&\pvec{\varphi_R}{\varphi_I}\ ,\quad \Shat=\pmat{U_R}{-U_I}{U_I}{U_R}\ ,
\end{align}
with subscripts $R$, $I$ indicating the real and imaginary parts and $\widehat{S}$ the real step evolution operator. We actually will work here with Majorana fermions and discretized time. In this case the wave function is real from the beginning and the evolution operator~$\Shat$ is an orthogonal matrix. (By introducing an appropriate complex structure (inverting eq.~\eqref{I1}) Majorana fermions in four dimensions are equivalent to Weyl fermions in a complex setting~\cite{CWMW1,CWMW2}.)

An automaton obtains if the real step evolution operator $\Shat(t)$ is a \qq{unique jump matrix}, which contains in each row and column precisely one element $\pm1$, with all other elements vanishing. In this case the step evolution operator maps each configuration of occupation numbers to precisely one new configuration. The unique jump property of $\Shat(t)$ is a particular restriction on the dynamics of the fermion system -- generic fermion systems are not described by an automaton. Nevertheless, rather rich families of two-dimensional fermionic quantum field theories with interactions have been found to admit a description as automata~\cite{FPCA}. They include models with Lorentz symmetry in the naive continuum limit, as discrete versions of Thirring~\cite{THI,KLA,AAR,FAIV} or Gross-Neveu~\cite{GN,WWE,RSHA,RWP,SZKSR} type models, models with local non-abelian gauge symmetries or discrete versions of spinor gravity. The true continuum limit for these models with the associated renormalization program has not yet been investigated, however, except for the trivial cases of free fermions.

In this note we ask the question if some interesting four-dimensional fermionic quantum field theories with Lorentz symmetry are equivalent to cellular automata. Instead of a wide general discussion we concentrate for a positive answer on a discrete model of spinor gravity in four dimensions. The local Lorentz symmetry is exact in the discrete version, while the diffeomorphism symmetry of general coordinate transformations is realized only in the naive continuum limit and assumed to emerge in the true continuum limit. 

The true continuum limit and its relation to the naive continuum limit are a central issue for the present paper. Establishing fully this limit for probabilistic cellular automata with interactions has to wait for the development of more sophisticated methods as functional renormalization for cellular automata or dedicated numerical simulations for the associated generalized Ising models. We develop here first steps that reveal the crucial importance of coarse graining and the new qualitative features associated to it.

Our method proceeds as follows. We start from the Grassmann functional integral for a fermionic quantum field theory formulated on a discrete space-time lattice. For this discrete fermion model we compute the step evolution operator according to the general methods of refs.~\cite{CWPCA, CWNEW, FPCA}. The naive continuum limit is defined by taking the limit of a vanishing lattice distance $\eps$ in the microscopic action. This neglects the renormalization flow of couplings and the question if symmetry violating interactions correspond to \qq{irrelevant operators} as suggested by the naive continuum limit. At a later stage functional methods for Grassmann integrals should be employed to establish the true continuum limit.

We will discuss several general building blocks for constructing discrete fermionic quantum field theories which are equivalent to probabilistic cellular automata and realize important symmetries in the naive continuum limit. This concerns a general strategy for implementing exact local gauge symmetries for cellular automata, the realization of a non-trivial evolution with the unique jump property by using an updating with shifted cells, the implementation of continuous or discrete symmetries in the naive continuum limit by sequences of different step evolution operators, and the emergence of diffeomorphism invariance in the naive continuum limit without the use of an explicit metric or vierbein. We explain all these techniques by a focus on a particular model for spinor gravity.

In sect.~\ref{Sect. II} we briefly introduce spinor gravity and discuss the properties that an associated automaton has to obey. Sect.~\ref{Sect. III} discusses simple automata with local Lorentz symmetry corresponding to fermionic actions which are quadratic in the invariants. The naive continuum limit does not yet account for propagating particles for this type of automata, however. In sect.~\ref{Sect. IV} we introduce interactions, as reflected by terms with four or more Lorentz-invariants in the fermionic action. The naive continuum limit admits now a non-trivial evolution. Sect.~\ref{Sect. V} introduces alternating updatings which lead to a rotation invariant naive continuum limit. In this limit diffeomorphism invariance is realized as well. As a result, the naive continuum limit is the spinor gravity model of sect.~\ref{Sect. II}. We draw conclusions in sect.~\ref{Sect. VI}.

\section{Spinor gravity automaton\label{Sect. II}}

In this section we briefly describe a particular continuum model for spinor gravity. Our spinor gravity automaton will be associated to a particular discretization of this model which realizes a unique jump step evolution operator~$\widehat{S}$. We establish the properties that the discrete Grassmann functional integral has to obey in order to implement the unique jump property of $\widehat{S}$.

\subsection*{Spinor gravity model}
\medskip

We investigate four-dimensional spinor gravity models~\cite{CWLSG,CWSGDI,SG3,SG4} based on Majorana fermions with eight flavors. The quantum field theory is formulated as a functional integral over Grassmann variables $\psi_\beta^A(x)$, with spinor index $\beta=1\dots4$, and flavor index $A=1\dots8$, $x=(t,\vec{x})$. The totally antisymmetric product of four Majorana spinors,
\begin{align}
\label{I2}
\chi^A(x)=&\frac1{24}\eps^{\eta\delta\alpha\beta}\psi_\eta^A(x)\psi_\delta^A(x)\psi_\alpha^A(x)\psi_\beta^A(x)\nn\\
=&\,\psi_1^A(x)\psi_2^A(x)\psi_3^A(x)\psi_4^A(x)\ ,
\end{align}
is invariant under local Lorentz transformations 
\bel{2A}
\delta\psi_\beta^A-\frac{1}{2}\varepsilon_{np}(x)\gl\Sigma^{np}\gr_{\beta\eta}\psi_\eta^A  .
\ee
Summation over repeated indices is always implied, with the exception of $A$ in the present case.
An action that is invariant under local Lorentz transformations as well as diffeomorphisms can be constructed as
\begin{align}
\label{I3}
S=\int_x& G_{ABCD,EFGH}\eps^{\mu\nu\rho\sigma}\nn\\
&\times\partial_\mu\chi^A\partial_\nu\chi^B\partial_\rho\chi^C\partial_\sigma\chi^D\chi^E\chi^F\chi^G\chi^H\ .
\end{align}
The invariance with respect to general coordinate transformations is achieved by contracting four derivatives with the totally antisymmetric tensor $\eps^{\mu\nu\rho\sigma}$. The coefficients $G$ define the particular model, where we note the total antisymmetry in the first four flavor indices, and symmetry in the last four indices. We will find a particular choice of $G$ and a particular discretization on a lattice for which this model is equivalent to a cellular automaton.

The action~\eqref{I3} shares all continuous symmetries of a fundamental theory of quantum gravity. The idea of spinor gravity is that appropriate collective fields based on products of spinor fields $\psi_\beta^A(x)$ acquire non-vanishing vacuum expectation values corresponding to a vierbein and associated metric. 
Indeed, a quantity with the homogeneous transformation properties of a vierbein (vector with respect to Lorentz symmetry and vector with respect to diffeomorphisms) can be formed from Grassmann variables (no sum over $A$ in eqs.~\eqref{3A},~\eqref{3B},~\eqref{3E}),
\bel{3A}
e\mb{}^{Am}=\psihat^A\gamma^{m}\partial\mb\psi^A=\psihat^A_{\gamma}(\gamma^{m})_{\gamma\delta}\partial_{\mu}\psi^A_{\delta}\ .
\ee
Here
\bel{3B}
\psihat^A_{\gamma}=-\frac{1}{24}\varepsilon_{\gamma\delta\alpha\beta}\psi^A_{\delta}\psi^A_{\alpha}\psi^A_{\beta}\;,\quad\chi^A=\psihat^A\psi^A=\psihat^A_{\gamma}\psi^A_{\gamma}  \ ,
\ee
and the real Dirac matrices in a Majorana basis are given by \cite{CPMW,Wetterich:2011zi}, in terms of the Pauli matrices $\tau_k$,
\begin{align}\label{3C}
\gamma^{0}=\pmat{0}{\tau_1}{-\tau_1}{0}\; ,\quad 
\gamma^{1}=\pmat{-\tau_1}{0}{0}{\tau_1}\ , \nn\\
\gamma^{2}=\pmat{-\tau_3}{0}{0}{-\tau_3}\; ,\quad
\gamma^{3}=\pmat{0}{\tau_1}{\tau_1}{0}\ .
\end{align}
The possible inhomogeneous part, due to the derivative in eq.~\eqref{3A} acting on the transformation parameters $\varepsilon_{np}(x)$, cancels by virtue of the identity 
\bel{3D}
\psihat^A\gamma^{m}\Sigma^{np}\psi^A=0\ ,
\ee
where the generators of the Lorentz group obey
\bel{3E}
\Sigma^{np}=-\frac{1}{4}[\gamma^{n},\gamma^{p}]\ .
\ee

In the presence of suitable expectation values an expansion of the multi-fermion action~\eqref{I3} in quadratic order in $\psi$ can induce an effective kinetic term for single fermions. If the expectation values are invariant under a global Lorentz symmetry which combines the global subgroup of the local Lorentz transformations with appropriate general coordinate transformations, this kinetic term can generate a (generalized) Dirac equation for the propagation of single fermions. The effective fermion kinetic term will become visible in a mean field approximation. Beyond this approximation it can be found in the quantum effective action which includes collective fields.

\subsection*{Step evolution operator for discrete fermionic quantum field theory}
\medskip

For a formulation of a fermionic quantum field theory on a discrete spacetime-lattice a general formulation for the extraction of the step evolution operator is available~\cite{CWFGI,CWPCA}.
This differs from other, more particular, fermionic descriptions of two-dimensional Ising models~\cite{PLECH,BER1,BER2,SAM,ITS,PLE1} or fermion-boson equivalence~\cite{FUR,NAO,COL,DNS}. Our formulation employs local factors $\K(t)$
\bel{ST1}
e^{-S}=e^{-\sum_t\cL(t)}=\prod_t\K(t)\ ,
\ee
which involve only Grassmann variables $\psi(t)$ and $\psi(\te)$ at neighboring times. A double expansion of $\K(t)$ in Grassmann basis elements $g_\rho(t)$ and $g_\tau(\te)$,
\bel{ST2}
\K(t)=g_\tau(\te)\Shat_{\tau\rho}(t)g'_\rho(t)\ ,
\ee
yields directly the elements of the step evolution operator $\Shat(t)$. The Grassmann basis elements $g_\rho(t)$ are products of Grassmann variables $\psi_\beta^A(t,\vec{x})$, where $g'_\rho$ is related to $g_\rho$ by a sign $\pm1$~\cite{CWPCA}.

For an intuitive interpretation the product $g_\tau(\te)g'_\rho(t)$ describes at $t$ a bit $n_\alpha(\vec x)=n_\beta^A(\vec{x})=1$ of type $\alpha=(\beta,A)$ at position $\vec{x}$ for every factor $\psi_\beta^A(t , \vec{x})$ present in $g'_\rho(t)$, while at $\te$ one has a bit $n_\delta^B(\vec{y})=1$ for every factor $\psi_\delta^B(\te , \vec{y})$ in $g_\tau(\te)$. The bits of all types for which no corresponding Grassmann variables appear in $g_\tau(\te)g'_\rho(t)$ take a zero value. Both $\rho$ and $\tau$ therefore specify bit-configurations. Then the non-zero elements of the unique jump step evolution operator $\Shat_{\tau\rho}(t)$ indicate for each bit configuration $\rho$ at $t$ a new bit configuration $\tau$ at $\te$. This specifies the updating of the automaton in the step from $t$ to $\te$. The uniqueness of the updating requires that each element $g'_\rho(t)$ in $\K(t)$ is multiplied by one single element $g_\tau(\te)$, with coefficient $\pm1$. The bits $n_\beta^A (\vec{x})=1$ can be identified with a fermionic particle of type $(\beta,A)$ present at the position $\vec{x}$, while $n_\beta^A(\vec{x})=0$ corresponds to a \qq{hole} of type $(\beta,A)$ or an absent fermionic particle at the position $\vec{x}$. The bits $n_\beta^A(\vec{x})=(1,0)$ equal the corresponding fermionic occupation numbers. The task consists in finding Grassmann expressions for $\cL(t)$ which realize the unique jump property for $\K(t)=\exp\l-\cL(t)\gr$.

Due to the modulo two property of Grassmann functional integrals~\cite{CWMW2} the identification~\eqref{ST2} holds only for odd $t$. For even $t$ the role of particles and holes is interchanged, and one employs a different, but analogue, formula for the relation between the local factor $\K(t)$ and the step evolution operator $\Shat(t)$~\cite{CWNEW}. We will focus here on models with particle-hole symmetry for which this complication is absent.

A bit-fermion map associates generalized Ising models and Grassmann functional integrals which lead to identical step evolution operators. Besides the identical evolution operators this bit-fermion map extends to wave-functions, operators for observables and expectation values~\cite{CWFGI}. We observe that the map from a Grassmann functional integral to an evolution sequence of bit configurations is unique, since $\K(t)$ defines $\Shat(t)$ uniquely. The opposite is not true, since a given bit sequence specifies the non-zero elements $\Shat_{\tau\rho}(t)$ only up to a sign. All choices of signs are equivalent, being related by local discrete gauge transformations~\cite{CWIT,CWQF}. We will choose signs such that the naive continuum limit of the discrete fermion models becomes simple. The bit-fermion map extends to arbitrary Grassmann functional integrals with time-local action~\eqref{ST1}, and to arbitrary step evolution operators. We focus here on the unique jump step evolution operators corresponding to automata.

\subsection*{Discrete spinor gravity model}
\medskip

The aim of this note is to find a discretization of the spinor gravity models~\eqref{I3}, such that an appropriate choice of $G$ yields the unique jump step evolution operator for an automaton. We group several microscopic cells corresponding to lattice points $\vec x$ into \qq{local cells}. These local cells are defined as non-overlapping cubes in a three-dimensional lattice. The updating of the bit-configuration in a local cell from $t$ to $\te$ depends only on the bit configuration in the same local cell. For the next updating step from $\te$ to $t+2\eps$ the positions of the local cells in the three-dimensional lattice are shifted, such that the bit configuration at $t+2\eps$ is influenced by the bit configurations in nine neighboring local cells at $t$. This locality property characterizes cellular automata. It induces the causal structure characteristic for local quantum field theories.

The task of finding a discrete model of the cellular automaton type remains rather complex, since we have to guarantee that every allowed bit or fermion configuration at $t$ is updated to precisely one allowed configuration at $\te$. The allowed configurations will be those for which at every site of the three-dimensional lattice the fermions can be grouped into a certain number of Lorentz invariants $\chi^A$. The total number of fermions therefore equals $0\ \text{mod}4$. The action~\eqref{I3} contains no terms with zero, one, two or three derivatives. It is odd under an arbitrary exchange of coordinates. The corresponding discrete symmetries will provide important guiding lines for the construction of a lattice action. We will realize these symmetries of the naive continuum limit by a sequence of $24$ different updatings which change the relative orientations in position and flavor space. Combining this sequence in a \qq{large updating step} by $\Delta t=24\eps$ the large updatings are then repeated.

\section{Automaton for free particles\label{Sect. III}}

This section discusses an automaton for ``free particles", in the sense that the fermionic action is quadratic in local Lorentz invariant combinations $\chi^A$. The particles are gauge invariant composites of fermions, somewhat similar to mesons. We first formulate a switch automaton in disconnected cells corresponding to hypercubes in a four-dimensional lattice. Propagation is introduced by shifting the cells in neighboring time layers. Comparison with the switch automaton without this shift reveals already important aspects of the naive continuum limit. At this level there is no standard propagation in the naive continuum limit.

\subsection*{Local updatings}
\medskip

\begin{figure}[h]
\includegraphics{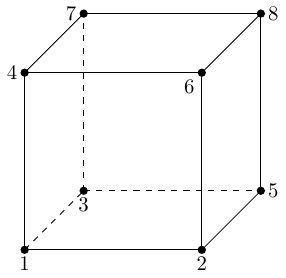}
\caption{Labeling of corners of cube.}
\label{fig:1}
\end{figure}

The updating from $t$ to $\te$ is done independently for each cell which corresponds to a cube with eight sites $u$ at the corners.
Our basic building blocks are the local Lorentz-invariants at every site,
\bel{A1}
\chi_u^{A}(t)=\psi_{u,1}^{A}(t)\psi_{u,2}^{A}(t)\psi_{u,3}^{A}(t)\psi_{u,4}^{A}(t)\ .
\ee
At every corner $u$ of a cube and for every flavor $A=1\dots N_f$ they are products of the four Grassmann variables $\psi_{u,\beta}^{A}(t)$. The index $u$ labels the positions at the corners of a cube as $(x,y,z)=1$, $(x+\eps,y,z)=2$, $(x,y+\eps,z)=3$, $(x,y,z+\eps)=4$, $(x+\eps,y+\eps,z)=5$, $(x+\eps,y,z+\eps)=6$, $(x,y+\eps,z+\eps)=7$, $(x+\eps,y+\eps,z+\eps)=8$, as shown in Fig.~\ref{fig:1}.
The local updating of the automaton is described by a four-dimensional hypercube which combines the cube at $t$ with the same cube at $\te$. The cubes constitute the local cells of the automaton. The state of the cube at $\te$ is determined  as a function of the state of the same cube at $t$. We place invariants $\chi^{A}$ at various positions in the hypercube. Denoting the center of the hypercube by $w\mo=(t+\frac\eps2,x+\frac\eps2,y+\frac\eps2,z+\frac\eps2)$, the positions of the corners of the hypercube are given by shifting each coordinate by $\pm\eps/2$. The signs for the different corners corresponding to our labels $u$ are listed in table~\ref{tab:1}.
\begin{table}

\centering
\renewcommand{\arraystretch}{1.2}

\begin{tabular}{|c|c|}
\hline
$\chi_1(t)$ & $----$\\
\hline
$\chi_2(t)$ & $-+--$\\
$\chi_3(t)$ & $--+-$\\
$\chi_4(t)$ & $---+$\\
\hline
$\chi_5(t)$ & $-++-$\\
$\chi_6(t)$ & $-+-+$\\
$\chi_7(t)$ & $--++$\\
\hline
$\chi_8(t)$ & $-+++$\\
\hline
\end{tabular}\quad\quad
\begin{tabular}{|c|c|}
\hline
$\chi_1(\te)$ & $+---$\\
\hline
$\chi_2(\te)$ & $++--$\\
$\chi_3(\te)$ & $+-+-$\\
$\chi_4(\te)$ & $+--+$\\
\hline
$\chi_5(\te)$ & $+++-$\\
$\chi_6(\te)$ & $++-+$\\
$\chi_7(\te)$ & $+-++$\\
\hline
$\chi_8(\te)$ & $++++$\\
\hline
\end{tabular}

\caption{Labeling of the positions of corners of the four-dimensional hypercube. With $w\mo$ the position of the center of the hypercube the locations are at $w\mo+s\mo\eps/2$, with signs $s\mo=(s^0,s^1,s^2,s^3)$ listed in the table.}
\label{tab:1}

\end{table}

Time-local factors $\K(t)$ are products of space-time-local factors $\cM(w)$
\bel{A2}
\K(t)=\prod_{x,y,z}\cM(w)=\prod_{x,y,z}\cM(t,x,y,z)\ ,
\ee
with $w=w(t,x,y,z)$. Here $\cM(w)$ is a function of the local Lorentz-invariants $\chi^{A}$ placed at different corners of the hypercube $w$. Thus $\K(t)$ involves Grassmann variables at $t$ and $\te$. The product over $x,y,z$ extends over a three-dimensional cubic lattice with lattice distance $2\eps$. Therefore two neighboring cubes at a given $t$ do not overlap and we can discuss the updating for each hypercube separately. A connection between different cubes or cells of the automaton will be established by shifting the positions of the hypercubes at the next time step from $\te$ to $t+2\eps$. We will express the space-time-local factors $\cM(w)$ in an exponential form
\bel{A3}
\cM(w)=\exp\gl-L(w)\gr\ .
\ee
The quantity $L(w)$ describes the elementary processes of the automaton, with $\cL (t)=\sum_{x,y,z} L(w)$.

\subsection*{Transport factors}
\medskip

Let us introduce transport factors involving one invariant at $t$ and one invariant at $\te$, for which we use the shorthands
\bel{A4}
\chi_u^{A}=\chi_u^{A}(t)\ ,\quad {\chi'}_u^{A}=\chi_u^{A}(\te)\ .
\ee
We define
\begin{align}
\label{A5}
T_{x\pm}^{AB}&=\chiu1A\chipu2B\pm\chiu2A\chipu1B+\chiu3A\chipu5B\pm\chiu5A\chipu3B\nn\\
&+\chiu4A\chipu6B\pm\chiu6A\chipu4B+\chiu7A\chipu8B\pm\chiu8A\chipu7B\ ,
\end{align}
\begin{align}
\label{A6}
T_{y\pm}^{AB}&=\chiu1A\chipu3B\pm\chiu3A\chipu1B+\chiu2A\chipu5B\pm\chiu5A\chipu2B\nn\\
&+\chiu4A\chipu7B\pm\chiu7A\chipu4B+\chiu6A\chipu8B\pm\chiu8A\chipu6B\ ,
\end{align}
\begin{align}
\label{A7}
T_{z\pm}^{AB}&=\chiu1A\chipu4B\pm\chiu4A\chipu1B+\chiu2A\chipu6B\pm\chiu6A\chipu2B\nn\\
&+\chiu3A\chipu7B\pm\chiu7A\chipu3B+\chiu5A\chipu8B\pm\chiu8A\chipu5B\ .
\end{align}
Without going into formal details, the interpretation of a term $\chi_u^A\chi_v^{\prime B}$ is simple. The four fermions $\chi_u^A$ are annihilated, and the four fermions forming $\chi_v^{\prime B}$ are created. Thus the transport factor $T_{x}^{AB}$ moves an invariant of flavor $A$, present at $t$ at position $(x,y,z)$, $(x,y+\eps,z)$, $(x,y,z+\eps)$ or $(x,y+\eps,z+\eps)$ (positions 1,3,4,7) to new positions with $x$ shifted to $x+\eps$. At $\te$ the corresponding positions are $(x+\eps,y,z)$, $(x+\eps,y+\eps,z)$, $(x+\eps,y,z+\eps)$ or $(x+\eps,y+\eps,z+\eps)$ (positions 2,5,6,8) respectively. Similarly, invariants at $x+\eps$ are moved to $x$. Simultaneously, the flavor changes from $A$ to $B$. Thus $T_x^{AB}$ describes a simultaneous switch in position and flavor. The transport factors $T_y$ and $T_z$ describe similar switches, with position changes now in the $y$- or $z$-directions. The position changes can be viewed as appropriate reflections within the cube. We also consider the flavor change without motion
\bel{A8}
T_0^{AB}=\sum_{u=1}^{8}\chiu{u}A\chipu{u}B\ .
\ee
A given invariant stays at its position and only changes its flavor from $A$ to $B$.

We can combine space reflections with flavor rotations by defining
\begin{align}
\label{A9}
\Ttil_x^{AB}&=T_{x-}^{AB}-T_{x-}^{BA}\ ,\quad &\Ttil_y^{AB}=T_{y-}^{AB}-T_{y-}^{BA}\ ,\nn\\
\Ttil_z^{AB}&=T_{z-}^{AB}-T_{z-}^{BA}\ ,\quad &\Ttil_0^{AB}=T_0^{AB}-T_0^{BA}\ .
\end{align}
The flavor $A$ is rotated to $B$, while the flavor $B$ is rotated to $A$ with an additional minus sign. This corresponds to a $\pi/2$-rotation in flavor space.

\subsection*{Switch automaton}
\medskip

Let us consider eight flavors, $A,B=1\dots8$. A simple switch automaton can be defined by
\bel{A10}
-L_{0}=\Ttil_x^{12}+\Ttil_y^{34}+\Ttil_z^{56}+\Ttil_0^{78}\ .
\ee
The fundamental process acts separately in the flavor subspaces $(1,2)$, $(3,4)$, $(5,6)$ and $(7,8)$. We can therefore write
\begin{align}
\label{A11}
\cM_{0}&=\exp\gl-L_{0}\gr\nn\\
&=\exp\gl\Ttil_x^{12}\gr\exp\gl\Ttil_y^{34}\gr\exp\gl\Ttil_z^{56}\gr\exp\gl\Ttil_0^{78}\gr\ .
\end{align}
We want to show that $\cM_0$ generates a unique jump step evolution operator in the space of Lorentz-invariant four fermion states $\chi_u^A$ within the cube $w$. It describes a unique rule how an arbitrary number of invariants, situated on arbitrary corners of the cube, is mapped to a new configuration of invariants in the cube. This updating map conserves the number of invariants in the cube. We can make the proof for each of the flavor subspaces independently.

Expanding the first factor
\bel{A12}
\exp\gl\Ttil_x^{12}\gr=1+\Ttil_x^{12}+\frac12\gl\Ttil_x^{12}\gr^2+\dots+\frac1{16!}\gl\Ttil_x^{12}\gr^{16}\ ,
\ee
the series terminates since $\gl\Ttil_x^{12}\gr^{17}=0$. This is due to the property $\gl\chiu uA\gr^2=0$, which is a direct consequence of the Grassmann property $\gl\psi_{u,\gamma}^{A}\gr^2=0$. In order sixteen the product $\gl\Ttil_x^{12}\gr^{16}$ contains one invariant $\chiu u1(t)$ and one invariant $\chiu u2(t)$ at every site $u$. The same happens at $\te$. The product of sixteen sums in $\gl\Ttil_x^{12}\gr^{16}$ involves $16!$ possibilities to combine these terms from the various factors. For each term the overall factor equals one, such that the factor $1/(16!)$ from the expansion of the exponential is compensated by $16!$ equal contributions. We identify
\begin{align}
\label{A13}
\frac{1}{16!}\gl\Ttil_x^{12}\gr^{16}=&\prod_{u=1}^{8}\chiu u1(t)\chiu u2(t)\chiu u1(\te)\chiu u2(\te)\nn\\
=&\prod_{\gamma=1}^{4}\prod_{u=1}^{8}\prod_{A=1}^{2}\psi_{u,\gamma}^{A}(t)\psi_{u,\gamma}^{A}(\te)\ ,
\end{align}
with the map from the totally filled (empty) state at $t$ to the totally filled (empty) state at $\te$.
Similarly, the first factor $1$ in the expansion of the exponential~\eqref{A12} maps the totally empty (filled) state to itself. The association of given Grassmann elements with particles or holes, or filled and empty, changes with  time due to the modulo two property of the Grassmann functional integral~\cite{CWPCA}. The corresponding two terms in the series~\eqref{A12} define a unit step evolution operator in the sectors of zero and sixteen invariants.

The elementary process $-L_{0}$ accounts for a reflection in position space accompanied by a flavor rotation for a single Lorentz-invariant. Within the restricted flavor space $(1,2)$ the \qq{switch factor} $\Ttil_x^{12}$ is a sum of $16$ terms, each relevant for the updating of a state with precisely one invariant of flavor $1$ or $2$, positioned at an arbitrary site $u$. The updating rule to a new state with only one invariant is unique, as appropriate for an automaton. For example, flavor $2$ at position $5$ is mapped to flavor $1$ at position $3$. Thus $-L_0$ describes a unique jump step evolution operator in the subsector with a single invariant.

The term $\gl\Ttil_x^{12}\gr^2/2$ in the expansion~\eqref{A12} describes the updating for states with two invariants ($8$ fermions). Due to $\gl\chiu uA\gr^2=0$ the two invariants either carry different flavor, or are positioned at different sites, or both. Every possible combination $\chiu uA(t)\chiu vB(t)$ appears in the product of the two sums. There are two possibilities with the same sign compensating the factor $1/2$ in the series~\eqref{A12}. The overall coefficient of this term is therefore $\pm1$. Every pair of invariants $\chiu uA(t)\chiu vB(t)$ is multiplied by a unique factor $\chiu{u'}{A'}(\te)\chiu{v'}{B'}(\te)$, corresponding to the unique updating rule for the automaton. A similar discussion applies to the term $\gl\Ttil_x^{12}\gr^3/(3!)$ which accounts for the unique updating of states with three different invariants in the cube. This generalizes to the higher terms in the expansion.

In conclusion, $\exp\gl\Ttil_x^{12}\gr$ describes the updating rule for an arbitrary number of different invariants of flavor $1$ or $2$ in the cell. Correspondingly, the factor $\exp\gl\Ttil_y^{34}\gr$ accounts for the unique updating of an arbitrary number of different invariants of flavor $3$ and $4$, and similarly for $\exp\gl\Ttil_z^{56}\gr$ for the flavors $5$ and $6$, and for $\exp\gl\Ttil_0^{78}\gr$ for the flavors $7$ and $8$. Taking things together the factor $\exp\gl-L_{0}\gr$ describes the unique updating for an arbitrary number of different invariants in the cell. The maximal total number of invariants (for the totally filled or empty state) amounts to $8N_f=64$. The expansion of $\cM_0$ in the occupation number basis generates a unique jump step evolution operator for invariants in a given cube of block or local cell. The combined local factor~\eqref{A2} generates an invertible updating rule for an arbitrary number of invariants in the space-lattice, and a corresponding unique jump evolution operator $\Shat(t)$.

\subsection*{Naive continuum limit}
\medskip

A naive continuum limit for the switch automaton can be constructed by introducing lattice derivatives which turn into partial derivatives in the limit $\eps\to0$. This limit requires a sufficiently smooth probabilistic information, i.e. smoothness of the wave function $q_\tau(t)$. The naive continuum limit simply keeps only the leading order in an expansion in $\eps$. This does not take into account fluctuation effects which typically induce running couplings which may or may not be compatible with the naive continuum limit. This issue is in analogy to the discretization of other quantum field theories as lattice gauge theories.

We express the invariants at the different corners of the cube in terms of an average field $\chibar$ and lattice derivatives as
\begin{align}
\label{C1}
\chi_1(t)=&\chibar-\frac\eps2\gl\partial_t+\partial_x+\partial_y+\partial_z\gr\chi+\dots\nn\\
\chi_2(t)=&\chibar-\frac\eps2\gl\partial_t-\partial_x+\partial_y+\partial_z\gr\chi+\dots\nn\\
\chi_3(t)=&\chibar-\frac\eps2\gl\partial_t+\partial_x-\partial_y+\partial_z\gr\chi+\dots\nn\\
\chi_4(t)=&\chibar-\frac\eps2\gl\partial_t+\partial_x+\partial_y-\partial_z\gr\chi+\dots\nn\\
\chi_5(t)=&\chibar-\frac\eps2\gl\partial_t-\partial_x-\partial_y+\partial_z\gr\chi+\dots\nn\\
\chi_6(t)=&\chibar-\frac\eps2\gl\partial_t-\partial_x+\partial_y-\partial_z\gr\chi+\dots\nn\\
\chi_7(t)=&\chibar-\frac\eps2\gl\partial_t+\partial_x-\partial_y-\partial_z\gr\chi+\dots\nn\\
\chi_8(t)=&\chibar-\frac\eps2\gl\partial_t-\partial_x-\partial_y-\partial_z\gr\chi+\dots\ .
\end{align}
Similar relations hold for $\chi_w(\te)$, with $\partial_t$ replaced by $-\partial_t$ in eq.~\eqref{C1}. This yields relations of the type
\begin{align}
\label{C2}
\chi_8(t)-\chi_1(t)=&\,\eps\gl\partial_t+\partial_x+\partial_y+\partial_z\gr\chi\ ,\nn\\
\frac12\gl\chi_8(t)+\chi_1(t)\gr=&\,\chibar+\dots\ .
\end{align}
The dots in eqs.~\eqref{C1},~\eqref{C2} denote corrections $\sim\eps^2$ with two or more derivatives. More precisely, we define
\bel{C3}
\eps\partial_t\chi=\frac18\sum_{u=1}^{8}\gl\chi_u(\te)-\chi_u(t)\gr\ ,
\ee
\bel{C4}
\chibar=\frac1{16}\sum_{u=1}^{8}\gl\chi_u(\te)+\chi_u(t)\gr\ ,
\ee
and
\begin{align}
\label{C5}
\eps\partial_x\chi=&\frac18\Big\{\big[\chi_2(t)+\chi_5(t)+\chi_6(t)+\chi_8(t)\nn\\
&\quad-\chi_1(t)-\chi_3(t)-\chi_4(t)-\chi_7(t)\big]+\gl t\leftrightarrow\te\gr\Big\}\nn\\
=&\frac18\Big\{\big[\chi(t,x+\eps,y,z)+\chi(t,x+\eps,y+\eps,z)\nn\\
&\quad+\chi(t,x+\eps,y,z+\eps)+\chi(t,x+\eps,y+\eps,z+\eps)\nn\\
&\quad-\chi(t,x,y,z)-\chi(t,x,y+\eps,z)-\chi(t,x,y,z+\eps)\nn\\
&\quad-\chi(t,x,y+\eps,z+\eps)\big]+\gl t\leftrightarrow\te\gr\Big\}\ ,
\end{align}
with similar definitions for $\partial_y\chi$ and $\partial_z\chi$. This partially specifies the dots in eqs.~\eqref{C1},~\eqref{C2}.

Expressing the transport factors~\eqref{A5}-~\eqref{A8} in terms of lattice derivatives one finds
\begin{align}
\label{C6}
T_{x-}^{AB}&=4\eps\gl\chi^{A}\partial_x\chi^B-\chi^B\partial_x\chi^{A}\gr\nn\\
&-2\eps^2\gl\partial_t\chi^{A}\partial_x\chi^B+\partial_t\chi^B\partial_x\chi^{A}\gr+\dots
\end{align}
and
\bel{C7}
T_{x+}^{AB}=8\chi^{A}\chi^B+4\eps\gl\chi^{A}\partial_t\chi^B-\chi^B\partial_t\chi^{A}\gr+\dots\ .
\ee
For $T_{y\pm}^{AB}$ and $T_{z\pm}^{AB}$ one obtains similar expressions, with $\partial_x$ replaced by $\partial_y$ or $\partial_z$, respectively.
The first two terms in an expansion of $T_0^{AB}$ read
\bel{C8}
T_0^{AB}=8\chi^{A}\chi^B+4\eps\gl\chi^{A}\partial_t\chi^B-\chi^B\partial_t\chi^A\gr+\dots\ .
\ee
It differs from $T_{x+}^{AB}$, $T_{y+}^{AB}$, $T_{z+}^{AB}$ only in the order $\eps^2$. The combinations~\eqref{A9} are antisymmetric in flavor space, such that
\begin{align}
\label{C9}
\Ttil_x^{12}&=8\eps\gl\chi^1\partial_x\chi^2-\chi^2\partial_x\chi^1\gr\ ,\nn\\
\Ttil_y^{34}&=8\eps\gl\chi^3\partial_y\chi^4-\chi^4\partial_y\chi^3\gr\ ,\nn\\
\Ttil_z^{56}&=8\eps\gl\chi^5\partial_z\chi^6-\chi^6\partial_z\chi^5\gr\ ,\nn\\
\Ttil_0^{78}&=8\eps\gl\chi^7\partial_t\chi^8-\chi^8\partial_t\chi^7\gr\ .
\end{align}

The naive continuum limit for the time-local factor of the switch automaton~\eqref{A10} reads
\bel{22A}
\Ktil(t)=\exp\big\{8\eps\sum_{xyz}\Ttil(t,x,y,z)\big\}\ ,
\ee
with fermionic action
\bel{34A}
S=-8\varepsilon\sum_{t,x,y,z} \tilde{T}(t,x,y,z)
\end{equation}
and
\begin{align}
\label{22B}
\Ttil&=\chi^1\partial_x\chi^2-\chi^2\partial_x\chi^1+\chi^3\partial_y\chi^4-\chi^4\partial_y\chi^3\nn\\
&+\chi^5\partial_z\chi^6-\chi^6\partial_z\chi^5+\chi^7\partial_t\chi^8-\chi^8\partial_t\chi^7\ .
\end{align}
Here the \qq{continuum invariants} $\chi^A\equiv\chi^A(t,x,y,z)$ are related to the local Lorentz invariants in the discrete formulation by eqs.~\eqref{C3}-~\eqref{C5}, with $\chi^A$ identified with $\chibar^A$.
The action is quadratic in the invariants. This holds both for the discrete version and the naive continuum limit.

If we keep in the action~\eqref{34A} the discrete version of $\Ttil$,
\bel{36A}
8\eps\Ttil=\Ttil_x^{12}+\Ttil_y^{34}+\Ttil_z^{56}+\Ttil_0^{78}\ ,
\ee
the four-dimensional fermionic quantum field theory constitutes an example for a probabilistic automaton. The ``quadratic action" does not describe a standard propagation of particles for increasing time, however. The only time-dependence is the oscillation between the flavors 7 and 8. Since different flavor sectors are not connected the time evolution is not transported to the other sectors. Inspection of the action~\eqref{I3} for spinor gravity indicates that it may be composed of four factors of the type $\Ttil_j^{AB}$. This connects the different flavors. We will indeed construct a discrete spinor gravity model with this structure. For this reason we discuss a few more properties of the switch automaton.

\subsection*{Discrete transformations and symmetries}
\medskip

Eq.~\eqref{C9} can be understood in terms of discrete transformations as reflections and lattice rotations. These discrete transformations in the hypercube are an important construction principle. Instead of changing coordinates we realize discrete transformations as discrete transformations on the local Lorentz-invariants $\chi_u$, or the corresponding Grassmann variables $\psi_{u,\beta}$. We list various reflections in table~\ref{tab:2}, and some of the rotations in table~\ref{tab:3}. These discrete transformations are flavor blind, i.e. they act in the same way on all flavors.
\begin{table}

\vspace{10pt}
\centering
\renewcommand{\arraystretch}{1.2}

\begin{tabular}{|c|c|l|}
\hline
$T$ & $t\leftrightarrow-t$ & $\chi_u\leftrightarrow\chi'_u$\\
$P_x$ & $x\leftrightarrow-x$ & $\chi_1\leftrightarrow\chi_2$, $\chi_3\leftrightarrow\chi_5$, $\chi_4\leftrightarrow\chi_6$, $\chi_7\leftrightarrow\chi_8$\\
$P_y$ & $y\leftrightarrow-y$ & $\chi_1\leftrightarrow\chi_3$, $\chi_2\leftrightarrow\chi_5$, $\chi_4\leftrightarrow\chi_7$, $\chi_6\leftrightarrow\chi_8$\\
$P_z$ & $z\leftrightarrow-z$ & $\chi_1\leftrightarrow\chi_4$, $\chi_2\leftrightarrow\chi_6$, $\chi_3\leftrightarrow\chi_7$, $\chi_5\leftrightarrow\chi_8$\\
$P$ & $P_xP_yP_z$ & $\chi_u\leftrightarrow\chi_{9-u}$\\
\hline
$S_{xy}$ & $x\leftrightarrow y$ & $\chi_2\leftrightarrow\chi_3$, $\chi_6\leftrightarrow\chi_7$\\
$S_{yz}$ & $y\leftrightarrow z$ & $\chi_3\leftrightarrow\chi_4$, $\chi_5\leftrightarrow\chi_6$\\
$S_{zx}$ & $z\leftrightarrow x$ & $\chi_2\leftrightarrow\chi_4$, $\chi_5\leftrightarrow\chi_7$\\
\hline
$S_{tx}$ & $t\leftrightarrow x$ & $\chi_2\leftrightarrow\chi'_1$, $\chi_5\leftrightarrow\chi'_3$, $\chi_6\leftrightarrow\chi'_4$, $\chi_8\leftrightarrow\chi'_7$\\
$S_{ty}$ & $t\leftrightarrow y$ & $\chi_3\leftrightarrow\chi'_1$, $\chi_5\leftrightarrow\chi'_2$, $\chi_7\leftrightarrow\chi'_4$, $\chi_8\leftrightarrow\chi'_6$\\
$S_{tz}$ & $t\leftrightarrow z$ & $\chi_4\leftrightarrow\chi'_1$, $\chi_6\leftrightarrow\chi'_2$, $\chi_7\leftrightarrow\chi'_3$, $\chi_8\leftrightarrow\chi'_5$\\
\hline
\end{tabular}
\caption{Action of discrete reflection symmetries. For the symmetries $P_x$, $P_y$, $P_z$, $S_{xy}$, $S_{yz}$, $S_{zx}$ the bilinears $\chi'_u$ transform in the same way as $\chi_u$. Invariant bilinears are not listed.}
\label{tab:2}

\end{table}
\begin{table}

\vspace{10pt}
\centering
\renewcommand{\arraystretch}{1.2}

\begin{tabular}{|c|c|l|}
\hline
$R_{xy}$ & $\chi_1\to\chi_3\to\chi_5\to\chi_2\to\chi_1$ & $\chi_4\to\chi_7\to\chi_8\to\chi_6\to\chi_4$\\
$R_{yz}$ & $\chi_1\to\chi_4\to\chi_7\to\chi_3\to\chi_1$ & $\chi_2\to\chi_6\to\chi_8\to\chi_5\to\chi_2$\\
$R_{zx}$ & $\chi_1\to\chi_2\to\chi_6\to\chi_4\to\chi_1$ & $\chi_3\to\chi_5\to\chi_8\to\chi_7\to\chi_3$\\
\hline
\end{tabular}
\caption{Rotations in the $xy$, $yz$ and $zx$ planes. The transformation of $\chi'_w$ is the same as for $\chi_u$.}
\label{tab:3}

\end{table}

The factor $T_{x-}^{12}$ or $\Ttil_x^{12}$, abbreviated here by $T_x$, changes sign under the reflection of the $x$-coordinate $P_x$, while it is invariant under the two other coordinate reflections $P_y$ and $P_z$. Similarly, $T_y$ is odd under $P_y$ and $T_z$ is odd under $P_z$,
\begin{alignat}{3}
\label{N11}
&P_x(T_x)=-T_x\ ,&\quad &P_x(T_y)=T_y\ ,&\quad &P_x(T_z)=T_z\ ,\nn\\
&P_y(T_x)=T_x\ ,&\quad &P_y(T_y)=-T_y\ ,&\quad &P_y(T_z)=T_z\ ,\nn\\
&P_z(T_x)=T_x\ ,&\quad &P_z(T_y)=T_y\ ,&\quad &P_z(T_z)=-T_z\ .
\end{alignat}
One finds for the diagonal reflections
\begin{align}
\label{N12}
&S_{xy}(T_x)=T_y\ ,\quad &S_{xy}(T_y)=T_x\ ,\quad &S_{xy}(T_z)=T_z\ ,\nn\\
&S_{yz}(T_x)=T_x\ ,\quad &S_{yz}(T_y)=T_z\ ,\quad &S_{yz}(T_z)=T_y\ ,\nn\\
&S_{zx}(T_x)=T_z\ ,\quad &S_{zx}(T_y)=T_y\ ,\quad &S_{zx}(T_z)=T_x\ ,
\end{align}
and $\pi/2$ rotations result in
\begin{alignat}{3}
\label{N13}
&R_{xy}(T_x)=-T_y\ ,&\quad &R_{xy}(T_y)=T_x\ ,&\quad &R_{xy}(T_z)=T_z\ ,\nn\\
&R_{yz}(T_x)=T_x\ ,&\quad &R_{yz}(T_y)=-T_z\ ,&\quad &R_{yz}(T_z)=T_y\ ,\nn\\
&R_{zx}(T_x)=T_z\ ,&\quad &R_{zx}(T_y)=T_y\ ,&\quad &R_{zx}(T_z)=-T_x\ .
\end{alignat}

On the level of the lattice derivatives $P_x$ inverts the sign of $\partial_x$ while keeping the sign of $\partial_y$ and $\partial_z$. This holds analogously for the other two reflections $P_y$ and $P_z$. The reflection $S_{xy}$ exchanges $\partial_x\leftrightarrow\partial_y$, keeping $\partial_z$ fixed. Finally, the rotation $R_{xy}$ transforms $\partial_x\to-\partial_y\to-\partial_x\to\partial_y\to\partial_x$. These transformation properties explain to a large extent eqs.~\eqref{C6}-~\eqref{C8}. We observe that in these expressions the first terms $\sim\eps$ are $T$-even, while the second terms $\sim\eps^2$ are $T$-odd. The factor $\Ttil_x^{12}$ is odd under a combination of the reflection $t\leftrightarrow\te$ with a flavor exchange $1\leftrightarrow2$. This holds analogously for $\Ttil_y^{34}$ and $\Ttil_z^{56}$.

\subsection*{Updating with shifted cells}
\medskip

The spacetime-local updating of a single cell associates to each configuration with an arbitrary number of invariants (up to $N_f = 8$ invariants for any given site), located at arbitrary corners of the cube, and carrying arbitrary flavors, a new such configuration. The updating involves reflections within the cube and flavor rotations. Employing the updating twice leads to an identity transformation up to signs. A sequence of updatings of the same given cell leads to a very simple, but somewhat boring automaton.

\begin{figure}[h]
\includegraphics[width=0.5\textwidth]{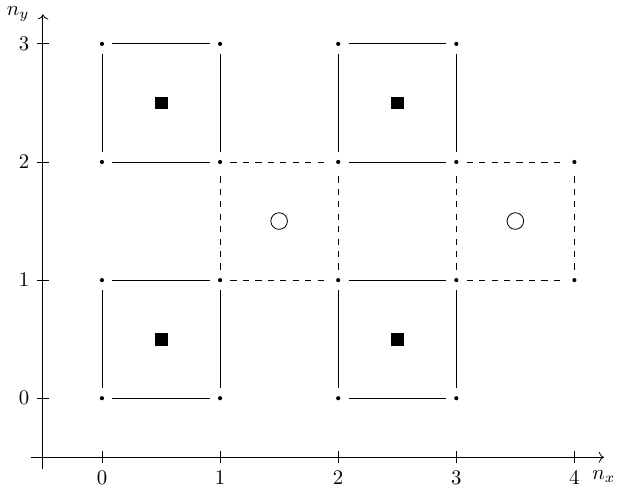}
\caption{Sequence of updatings with shifted cells for two space dimensions. The cells in the first updating step are the squares with centers indicated by the small filled solid squares. The sites of a given cell are the corners of a square, linked by the solid lines. The cells do not overlap. At the second updating step the cells are the squares with centers at the circles. Sites of a given cell are joined by the dashed lines. Again, the different cells do not overlap. Each site of a cell for the first step is found in a different cell for the second step.}
\label{fig:2}
\end{figure}

A richer structure obtains if we shift the position of the cells in consecutive updating steps. At the second updating step we will use a shifted lattice of cells. While for the updating from $t$ to $\te$ the centers of the hypercubes obey, with integers $n\mb$, 
\begin{align}
\label{US1}
w&=(w^0,w^1,w^2,w^3)\nn\\
&=\eps\gl2n_0+\frac12,2n_1+\frac12,2n_2+\frac12,2n_3+\frac12\gr\ ,
\end{align}
we employ for the updating from $\te$ to $t+2\eps$ centers of hypercubes located at
\bel{US2}
w=\gl2n_0+\frac32,2n_1+\frac32,2n_2+\frac32,2n_3+\frac32\gr\ .
\ee
The eight corners of the cube in the first updating step belong to eight different cubes in the second updating step. We illustrate this setting in Fig.~\ref{fig:2} for the case of two space dimensions. For example, the point $(\eps,0,\eps,\eps)$, which belongs to the hypercube $w=(\eps/2)(1,1,1,1)$ in the first updating step, will belong to the hypercube at $w=(\eps/2)(3,-1,3,3)$ in the second updating step. A different point in the same cell of the first updating, say $(\eps,\eps,0,0)$, belongs for the second step to the hypercube at $w=(\eps/2)(3,3,-1,-1)$. For the updating from $t+2\eps$ to $t+3\eps$ we use the same cells as for the one from $t$ to $\te$. The use of the same cells for $t$ and $t+2n_0\eps$ results in a sequence of alternating cell lattices.

The updating in the second step is again done for each cell separately. Combining two steps, however, the configuration of a given cell at $t+2\eps$ is influenced by the configurations of eight cells at $t$, and of $27$ cells at $t-\eps$. The cells that can influence a given cell $w$ at $t$ form the interior of a \qq{past light cone} of this cell. Similarly, the configuration of a given cell at $t$ influences the configurations of eight cells at $t+2\eps$. The cells that can be influenced by a given cell $w$ form the interior of the \qq{future light cone} of this cell. The linear extension of the light cone of $w$ grows $\sim\eps|\Delta t|$, with $\Delta t$ the time distance to the cell $w$ in the future or past. The updating with shifted cells therefore connects different parts of the lattice of cells. It exhibits a causal structure based on locality properties.

In principle, one may employ different updating rules for the individual cells at different steps in a sequence of updatings. We will indeed do so in later parts of this note. For the moment let us briefly discuss the automaton that follows if we apply the same updating rule~\eqref{A10} for the switch automaton for all updating steps. Consider at $t$ a single invariant of flavor $1$ located at $(x,y,z)=(0,0,0)$ or $u=1$. After the first updating we have at the site $(\eps,0,0)$ or $u=2$ an invariant of flavor $2$. In the second step this invariant at $(\eps,0,0)$ is situated at the site $u=7$ of the shifted cube. The updating moves it to the site $u=8$, now corresponding to the position $(2\eps,0,0)$. The flavor turns back to $1$, and the total result of the two steps is a transport in the positive $x$-direction by two units, keeping the flavor fixed. This extends to further steps, such that the invariant moves on a straight line in the $x$-direction.

The same behavior generalizes to invariants of flavor $1$ or $2$ which are situated at $t$ at the positions $1$, $3$, $4$, $7$. They move in the positive $x$-direction. Similarly, invariants positioned at $2$, $5$, $6$ or $8$ move in the negative $x$-direction. The \qq{right movers} and \qq{left movers} do not interfere. For example, one \qq{right mover} at position $1$ and one \qq{left mover} at position $2$ will be found for the second updating step in different cells and subsequently continue their motion, not influenced by the presence or absence of the other invariant. This generalizes to several invariants of flavor $1$ or $2$ in the cell at $t$. They move independently to the positive or negative $x$-direction, depending on their position at $t$.

The situation for invariants with flavors $3$ or $4$ is analogous. They move in the positive or negative $y$-direction, according to their position in the cube at $t$. For the flavors $5$ and $6$ the motion is in the positive or negative $z$-direction. Finally, the flavors $7$ and $8$ stay at their sites, alternating only in the type of flavor. In summary, this automaton describes a simple setting where flavors $1$, $2$ move on straight lines in the $x$-direction, flavors $3$, $4$ in the $y$-direction and flavors $5$, $6$ in the $z$-directions. The flavors $7$, $8$ are stationary. The flavor switches at alternating time steps. This automaton describes a type of motion of free particles, combined into Lorentz invariants, with direction of motion given by the flavor.

For the two alternating lattices of cells the translation of a given point to the location in the corresponding cell is simple. If the point $(x,y,z)$ corresponds to the position $1$ for the cells in the first lattice, it corresponds to the position $8$ for the cells of the shifted lattice. Similarly, the point $(x+\eps,y,z)$ has position $2$ in the cells of the first lattice, and position $7$ in the cells of the second lattice. Between the two lattices points in the cell are shifted according to maximal diagonals, or $u\leftrightarrow9-u$.

\subsection*{Continuum limit for shifted cells}
\medskip

The naive continuum limit for the automaton with shifted cells is again given by eqs~\eqref{34A},~\eqref{22B}. The shift of the cells is not visible in this limit. For the continuum limit there is no difference between the automaton with shifted cells and the automaton for which the same cells are used at every time step. On the other hand, we have discussed that on the discrete level the behavior of the two types of automata is rather different. This points to important properties, and also important limitations, of the naive continuum limit. 

The issue can be understood if we enhance the resolution of the continuum description. For the flavors 1 and 2 we may introduce an extra label $\eta=+1$ for those invariants that are located at $t$ at $(\overline{x}-\frac{\varepsilon}{2},\,y,\,z)$, and $\eta = -1$ if the position at $t$ is $(\overline{x}+\frac{\varepsilon}{2},\,y,\,z)$. For the automaton with shifted cells the invariants with $\eta=+1$ propagate towards increasing $x$, while the ones with $\eta=-1$ move to decreasing $x$. For a single time step one has an additional flavor switch, while for an even number of time steps the invariants $\eta=+1$ propagate as ``right -movers" in the $x$-direction without flavor change, while $\eta=-1$ denotes ``left-movers". It is possible to define a refined continuum limit for right-moving and left-moving particles. For this refined continuum limit the difference to the automaton without a shift is visible. Without the shift both types $\eta=\pm 1$ are stationary, not changing their position for an even number of time steps. 

The naive continuum limit employed here averages over the positions in the cell. It makes no difference between $\eta=+1$ and $\eta=-1$. For wave functions whose variation within a single cell can be neglected the probability of an invariant moving to the right or to the left is the same. The average position remains stationary, just as for the automaton without the shift. One expects that some form of dissipation should distinguish the two automata. This information is not kept, however, in the naive continuum limit. 

This simple discussion reveals a general property of the continuum limit. It involves some form of ``coarse graining" with the associated loss of information. On the one side, this loss of information is crucial to ensure a certain ``universality" of the continuum limit. On the other side, not all possible discretizations are expected to belong to the same ``universality class".

\section{Automaton with interactions\label{Sect. IV}}

In our setting interactions correspond to terms in the action for the Grassmann functional integral that involve terms with more than two invariants. This will bring us already close to the action~\eqref{I3} of the spinor gravity model where each term involves eight invariants. While it is straightforward to add interaction terms in $L$, the issue is to maintain the uniqueness of the updating which restricts the admitted properties of $\exp\gl-L\gr$.

\subsection*{Interaction}
\medskip

We use interactions in order to construct an automaton with local Lorentz invariance which leads to propagation of all invariants in the naive continuum limit. For this purpose we divide the eight flavors into four generations, each coming with two types. We use a double index notation, $A=(a,\sigma_a)$, $a=1\dots4$, $\sigma_a=\pm1$, where $a$ denotes the generation and $\sigma_a$ the associated type. Les us consider an automaton for which an interaction occurs if four invariants belonging all to different generations meet in a cube. We modify the updating rule such that for four invariants of different generations the updating still operates a reflection within the cube, but the flavor change is different. For this particular type of configuration the generation is maintained, but the flavor switch is now between types $\sigma_a=\pm1$. With flavor labels $A=(1,2,3,4,5,6,7,8)=(1+,2+,3+,4+,1-,2-,3-,4-)$ the new rule is implemented by an additional term in $L$
\bel{IN1}
-\Delta L=\Ttil_x^{15}\Ttil_y^{26}\Ttil_z^{37}\Ttil_0^{48}-\Ttil_x^{12}\Ttil_y^{34}\Ttil_z^{56}\Ttil_0^{78}\ .
\ee

Both therms in eq.~\eqref{IN1} are products of four sums, with $16$ terms in each sum. Each term in the total sum of $16^4=2^{16}$ terms contains precisely one invariant $\chiu{u}{1\pm}$, one invariant $\chiu{v}{2\pm}$, one invariant $\chiu{w}{3\pm}$ and one invariant $\chiu{z}{4\pm}$, such that all four generations are present in the cube, with either type $\sigma_a=+1$ or type $\sigma_a=-1$. The first term in eq.~\eqref{IN1} accounts for the simultaneous reflection for all four generations, where generation $1$ is reflected in the $x$-direction, generation $2$ in the $y$-direction, generation $3$ in the $z$-direction, while generation $4$ is stationary. Furthermore, the type is switched from $1+$ to $1-$ and vice versa. This is the scattering term.

The second term subtracts the part that would arise in the same sector from $\exp(-L_0)$, with $L_0$ corresponding to the switch automaton for free particles. Indeed, the cellular automaton property requires that each term in
\bel{IN2}
\exp\gl-L\gr=\exp\gl-L_0-\Delta L\gr\ ,
\ee
which contains a certain product of factors $\chiu uA(t)$, is multiplied by precisely one product of factors $\chiu uA(\te)$. The term $L_0^4 /24$ in the expansion of $\exp(-L_0)$ produces in the sector with four invariants of different generations at $t$ precisely the second term in eq.~\eqref{IN1}, with a plus sign. Subtracting it in $-\Delta L$ guarantees that a given factor, say $\chiu1{1+}(t)\chiu4{2-}(t)\chiu8{3-}(t)\chiu5{4+}(t)$ is precisely multiplied by $\chiu2{1-}(\te)\chiu7{2+}(\te)\chiu5{3+}(\te)\chiu5{4-}(\te)$, such that the updating is unique.

\subsection*{Uniqueness of updating}
\medskip

The uniqueness of the updating must be maintained for every possible combination of invariants in the cube. Furthermore, the updating map should be invertible and we want to implement particle-hole symmetry. This typically requires further \qq{correction terms} in $L$, which involve five or more invariants $\chi(t)$, and five or more invariants $\chi(\te)$. For example, on the level of five invariants one has
\bel{IN3}
\exp\big[-\gl(L_0+\Delta L+\Delta L'\gr\big]=-\frac1{5!}L_0^5+L_0\Delta L-\Delta L'+\dots\ .
\ee
We have to specify the updating rule for configurations where five invariants are present in the cube, two of them of the same generation, and the other three of the three different complementary generations. For example, we may postulate that in this situation no scattering occurs, fixing $\Delta L'=L_0\Delta L$. For a different rule in this sector $\Delta L'$ will be different. A simple possible updating rule could involve scattering only if the number of invariants of each generation in a cell is equal. In this case the reflections in the cube are accompanied by a switch of all types $\sigma$, keeping the generation unchanged. Specifying the updating rule fixes $\Delta L'$ up to possible signs of the different terms.

In the presence of the interaction term~\eqref{IN1} the naive continuum limit of the time-local factor becomes
\bel{IN4}
\Ktil(t)=\exp\bigg\{\sum_{xyz}\Big[8\eps\Ttil+(8\eps)^4\gl\Itil_1-\Itil_2\gr+\mathcal{O}(\eps^5)\Big]\bigg\}\ ,
\ee
with
\begin{align}
\label{IN5}
\Itil_1&=\gl\chi^1\partial_x\chi^5-\chi^5\partial_x\chi^1\gr\gl\chi^2\partial_y\chi^6-\chi^6\partial_y\chi^2\gr\nn\\
&\times\gl\chi^3\partial_z\chi^7-\chi^7\partial_z\chi^3\gr\gl\chi^4\partial_t\chi^8-\chi^8\partial_t\chi^4\gr\ ,
\end{align}
and
\begin{align}
\label{IN6}
\Itil_2&=\gl\chi^1\partial_x\chi^2-\chi^2\partial_x\chi^1\gr\gl\chi^3\partial_y\chi^4-\chi^4\partial_y\chi^3\gr\nn\\
&\times\gl\chi^5\partial_z\chi^6-\chi^6\partial_z\chi^5\gr\gl\chi^7\partial_t\chi^8-\chi^8\partial_t\chi^7\gr\ .
\end{align}
The corresponding term $\sim\Itil_1-\Itil_2$ in the fermionic action shows an important feature of the spinor gravity action~\eqref{I3}, namely the presence of four derivatives for the four different cartesian directions. The flavors are now all mixed. The field equations exhibit a time evolution which defines a type of propagation for all flavors.

The term $\Delta L'$ enters only in the order $\eps^5$ and will be neglected in the naive continuum limit. This highlights the limitations of the naive continuum limit. Different automata with different updating rules for five or more invariants in a cell entail different $\Delta L'$. They have nevertheless the same naive continuum limit. It is not clear a priori which updating rules lead to the universality class corresponding to the naive continuum limit, and which ones not. The universality class is typically characterized by continuous and discrete symmetries. It is not guaranteed that such symmetries are sufficient for a complete characterization of the possible universality classes.

The fermionic action for a simple updating rule may get rather complex. While the terms with a relatively low number of Grassmann variables reflect directly the updating rule and are rather simple, higher order terms as $\Delta L'$ are often needed to maintain the unique jump property of the automaton. In the naive continuum these higher order terms vanish. The true continuum limit has to investigate the renormalization flow of these higher order terms.

\section{Alternating updating and coarse \\  graining\label{Sect. V}}

Sequences of different updating steps are a powerful construction principle in order to realize symmetries in the naive continuum limit. Many symmetries lead in the naive continuum limit to a situation where a given configuration of occupation numbers or bits can evolve to several different configurations -- it has \qq{different options}. These options are typically related by symmetry transformations. The presence of different options is not compatible with a unique jump step evolution operator. For an automaton with a sequence of different updating steps each individual step may permit only one of the options. This preserves the automaton property, but violates the symmetry. Constructing the Hamiltonian by the evolution from $t$ to $t+\Delta t$ as in eq.~\eqref{INN8} amounts to some effective averaging or coarse graining. For coarse grained automata the unique jump property is typically lost. This permits us to realize the symmetry on a coarse grained level.

As an example, we may take lattice rotations by $\pi/2$ in some of the planes, $x\to y\to-x\to-y\to x$. A given single updating step may not respect the rotation symmetry as is the case for the switch automaton characterized by $L_0$ or the automaton with interaction specified by eq.~\eqref{IN3}. Rotating the updating for each step of the sequence by $\pi/2$ produces after four steps a rotation invariant naive continuum limit. In the naive continuum limit the contributions of the four updating steps are simply added without taking into account the ordering in time of the sequence. It remains an open issue if this approximation can be justified by the true continuum limit.

In eq.~\eqref{IN4} the term $\tilde{I}_{1}-\tilde{I}_{2}$ shows already some of the structures of spinor gravity. It does not yet show the characteristic contraction of four derivatives with the totally antisymmetric tensor $\varepsilon^{\mu\nu\rho\sigma}$. This contraction guarantees rotation symmetry. It is the basis of diffeomorphism invariance of the naive continuum limit. Furthermore, the local factor~\eqref{IN4} contains the term $\sim\tilde{T}$ which is not present for the spinor gravity model. We will introduce alternating updatings in order to ensure rotation invariance of the naive continuum limit and the absence of terms quadratic in the invariants. This highlights further the coarse graining involved for the continuum limit. 

\subsection*{Reflected updatings}
\medskip

The automaton with interactions defined by the local factor or associated fermionic action~\eqref{IN2}, with eqs.~\eqref{A10},~\eqref{IN1}, is invariant under certain lattice reflections and rotations combined with flavor rotations. It is not invariant under lattice reflections or rotations alone. As a consequence, the naive continuum limit~\eqref{IN4} is not rotation invariant. Rotation invariance of the naive continuum limit can be achieved if we change the updating rule of a given cell for consecutive steps of the updating. We will first consider a sequence of six steps. For each step the updating is given for individual cells, and we alternate the cell-lattice according to the previous discussion of shifted cells.

The first step is given by the rules~\eqref{A10},~\eqref{IN1}. In the second step we employ an updating that is obtained by acting with a reflection $S_{yz}$ on the previous one, together with an additional minus sign. This yields
\begin{align}
\label{Z1}
-L(\te)&=-\Ttil_x^{12}-\Ttil_y^{56}-\Ttil_z^{34}-\Ttil_0^{78}\nn\\
&-\Ttil_x^{15}\Ttil_y^{37}\Ttil_z^{26}\Ttil_0^{48}+\Ttil_x^{12}\Ttil_y^{56}\Ttil_z^{34}\Ttil_0^{78}\ .
\end{align}
The third step obtains from the second step by applying the reflection $-S_{zx}$, and the fourth step obtains from the previous one by applying the reflection $-S_{xy}$. For the fifth step we reflect the fourth step by $-S_{yz}$, and finally the sixth step is generated from the fifth step by the reflection $-S_{zx}$. Note that the first step obtains from the sixth step by $-S_{xy}$.

In the naive continuum limit the precise position of $t$ or $\te$ is no longer important. One finds
\begin{align}
\label{Z2}
-\gl L_0(t)+L_0(\te)\gr&=\Ttil_y^{34}+\Ttil_z^{56}-\Ttil_z^{34}-\Ttil_y^{56}\ ,\nn\\
-\gl L_0(t+2\eps)+L_0(t+3\eps)\gr&=\Ttil_x^{34}+\Ttil_y^{56}-\Ttil_y^{34}-\Ttil_x^{56}\ ,\nn\\
-\gl L_0(t+4\eps)+L_0(t+5\eps)\gr&=\Ttil_z^{34}+\Ttil_x^{56}-\Ttil_x^{34}-\Ttil_z^{56}\ .
\end{align}
The sum of all six terms vanishes. What remains are the interaction terms that take the form
\begin{align}
\label{Z3}
-\sum_{n_s=0}^{5}L(t+n_s\eps)=\eps^{ijk}\Big\{&\Ttil_i^{15}\Ttil_j^{26}\Ttil_k^{37}\Ttil_0^{48}\nn\\
-&\Ttil_i^{12}\Ttil_j^{34}\Ttil_k^{56}\Ttil_0^{78}\Big\}\ ,
\end{align}
with $i,j,k=1\dots3$, and $\eps^{ijk}$ the totally antisymmetric Levi-Civita tensor. (Here we employ $i=(1,2,3)=(x,y,z)$, and similar for $j$ and $k$.) Insertion of eq.~\eqref{C9} yields
\begin{align}
\label{Z4}
&-\sum_{n_s=0}^{5}L(t+n_s\eps)=(8\eps)^4\eps^{ijk}\nn\\
\times\Big\{&\gl\chi^1\partial_i\chi^5-\chi^5\partial_i\chi^1\gr\gl\chi^2\partial_j\chi^6-\chi^6\partial_j\chi^2\gr\nn\\
\times&\gl\chi^3\partial_k\chi^7-\chi^7\partial_k\chi^3\gr\gl\chi^4\partial_t\chi^8-\chi^8\partial_t\chi^4\gr\nn\\
-&\gl\chi^1\partial_i\chi^2-\chi^2\partial_i\chi^1\gr\gl\chi^3\partial_j\chi^4-\chi^4\partial_j\chi^3\gr\nn\\
\times&\gl\chi^5\partial_k\chi^6-\chi^6\partial_k\chi^5\gr\gl\chi^7\partial_t\chi^8-\chi^8\partial_t\chi^7\gr\Big\}\ .
\end{align}
Due to the contraction with the rotation invariant tensor $\eps^{ijk}$ the naive continuum limit is invariant under rotations in three-dimensional space, keeping the flavor fixed.

\subsection*{Diffusion}
\medskip

The vanishing continuum limit for $L_0$,
\bel{Z5}
-\sum_{n_s=0}^{5}L_0(t+n_s\eps)=0\ ,
\ee
does not imply that particles do not move in the absence of interactions. A single invariant of flavor $1$ or $2$, situated at $t$ at position $(x,y,z)=(0,0,0)$, will be found at $t+6\eps$ at the position $(2\eps,2\eps,2\eps)$, without changing its flavor. Similarly, an original position $(\eps,0,0)$ is moved to $(-\eps,2\eps,2\eps)$ at $t+6\eps$, while $(0,\eps,0)$ changes to $(2\eps,-\eps,2\eps)$ and $(0,0,\eps)$ to $(2\eps,2\eps,-\eps)$. The remaining four sites move as
\begin{alignat*}{2}
(\eps,0,\eps)&\to(-\eps,2\eps,-\eps)\ ,\quad (\eps,\eps,0)&&\to(-\eps,-\eps,2\eps)\ ,\\
(0,\eps,\eps)&\to(2\eps,-\eps,-\eps)\ ,\quad (\eps,\eps,\eps)&&\to(-\eps,-\eps,-\eps)\ .
\end{alignat*}

The average position for the eight corners of the cell at $t$ is given by $(\eps/2,\eps/2,\eps/2)$. This average position remains the same at $t+6\eps$. We conclude that the average position does not move for the sequence of six steps. There is, however, an effective diffusion since each site in the cell at $t$ is moved at $t+6\eps$ to a different cell. The naive continuum limit does not capture this diffusion. 

The situation for the other flavors is similar. We can follow the six updating steps without interactions by the rules summarized in table~\ref{tab:4}.
\begin{table}
\renewcommand{\arraystretch}{1.2}
\centering

\begin{tabular}{|c|c|c|c|c|}
\hline
& $1$, $2$ & $3$, $4$ & $5$, $6$ & sign\\
\hline
$t$ & $P_x$ & $P_y$ & $P_z$ & $+$\\
$t+\eps$ & $P_x$ & $P_z$ & $P_y$ & $-$\\
\hline
$t+2\eps$ & $P_z$ & $P_x$ & $P_y$ & $+$\\
$t+3\eps$ & $P_z$ & $P_y$ & $P_x$ & $-$\\
\hline
$t+4\eps$ & $P_y$ & $P_z$ & $P_x$ & $+$\\
$t+5\eps$ & $P_y$ & $P_x$ & $P_z$ & $-$\\
\hline
\end{tabular}
\caption{Updating for sequence of six steps for free particles. The first three columns denote the different flavors, with a flip of flavors at each step. We indicate the reflections within a cell for each step. For the lattice with shifted cells the direction of the reflection indicates the direction of motion in the corresponding positive or negative direction, depending on the starting point. The last column indicates the overall sign of $L_0(t+n_s\eps)$.}
\label{tab:4}
\end{table}
For the flavors $5$ and $6$ the positions move from $t$ to $t+6\eps$ as
\begin{alignat*}{2}
(0,0,0)&\to(-2\eps,-2\eps,2\eps)\ ,\quad &&(\eps,0,0)\to(3\eps,-2\eps,2\eps)\ ,\\
(0,\eps,0)&\to(-2\eps,3\eps,2\eps)\ ,\quad &&(0,0,\eps)\to(-2\eps,-2\eps,-\eps)\ ,\\
(\eps,\eps,0)&\to(3\eps,3\eps,2\eps)\ ,\quad &&(\eps,0,\eps)\to(3\eps,-2\eps,-\eps)\ ,\\
(0,\eps,\eps)&\to(-2\eps,3\eps,-\eps)\ ,\quad &&(\eps,\eps,\eps)\to(3\eps,3\eps,-\eps)\ .
\end{alignat*}
Again, the average position remains invariant. Correspondingly, the position change for the flavors $3$ and $4$ obeys
\begin{alignat*}{2}
(0,0,0)&\to(2\eps,2\eps,-2\eps)\ ,\quad &&(\eps,0,0)\to(-\eps,2\eps,-2\eps)\ ,\\
(0,\eps,0)&\to(2\eps,-\eps,-2\eps)\ ,\quad &&(0,0,\eps)\to(2\eps,2\eps,3\eps)\ ,\\
(\eps,\eps,0)&\to(-\eps,-\eps,-2\eps)\ ,\quad &&(\eps,0,\eps)\to(-\eps,2\eps,3\eps)\ ,\\
(0,\eps,\eps)&\to(2\eps,-\eps,3\eps)\ ,\quad &&(\eps,\eps,\eps)\to(-\eps,-\eps,3\eps)\ ,
\end{alignat*}
with conserved average position. Finally, the flavors $7$ and $8$ remain at their positions.

We conclude that the motion of invariants on individual sites is rather complex for the combined updating from $t$ to $t+6\eps$. Simplicity arises on the averaged level with no motion of the average position. It is this averaged property that is captured by the naive continuum limit with vanishing combined $L_0$. We recall that we have neglected terms with two or more derivatives acting on at least one invariant. These terms could describe part of the diffusive properties. We will discuss the justification of their neglection later.

What remains in the continuum limit are the interaction terms~\eqref{Z4}. They have no longer the unique jump property of an automaton. Indeed, if we leave the information on which particular site the invariants are located at $t$, and only keep the information about the number of invariants of given flavors in a cell, the average updating is no longer unique. For configurations with a fixed number of invariants with given flavors in a cell there remain several different possibilities or options for their evolution to different flavors and different cells. A coarse graining from the sites of a cell to the whole cell looses information. On the coarse grained level the automaton property is no longer present. This is a characteristic feature of any continuum limit.

The naive continuum limit can also be seen as a coarse graining over six time steps. In this coarse graining the order of the individual time steps is no longer resolved. Without the information on the precise order of the six time steps the automaton property is lost. Different orders of the updating steps lead to different possibilities for the detailed positions of the various invariants at $t+6\eps$. This feature is closely related to the fact that the different updating steps do not commute. This is directly visible in the different updating rules for different flavors on the level of invariants at resolved sites. The different updatings correspond to different orders of the individual steps. On the level of the invariant average position the order of the steps no longer matters. The sum of terms in the naive continuum limits~\eqref{Z3},~\eqref{Z4},~\eqref{Z5} does not retain any information about the order of the steps.

\subsection*{Flavor rotations and reflections}
\medskip

Instead of the coordinate reflections defining the six steps we can achieve the same automaton by rotations in flavor space. The propagation term $L_0$ is invariant under a combination of the reflection $S_{xy}$ and an exchange of the flavor pairs $(1,2)$ and $(3,4)$. The latter can be realized by the rotation in flavor space $F_{12}F_{24}$, where we define
\bel{B0A}
F_{AB}:\ \chi^A\to\chi^B\to-\chi^A\to-\chi^B\to\chi^A\ .
\ee
Combining appropriate rotations, as $F_{13}F_{42}$ or $F_{14}F_{23}$, with flavor reflections $G_{12}G_{78}$, one can realize the additional minus sign in our choice of the updating steps. Here we define the flavor reflection as
\bel{B0B}
G_{AB}:\ \chi^A\leftrightarrow\chi^B\ .
\ee
In other words, a change of the updating rule by the reflection $S_{xy}$ with an additional minus sign can equivalently be achieved by an appropriate rotation in flavor space. (We use here rotations for generalized rotations including reflections.) Similarly, one can replace the transformation $-S_{yz}$ by the flavor rotations $F_{35}F_{64}$ or $F_{36}F_{45}$, combined with $G_{12}G_{78}$, and $-S_{zx}$ by $F_{15}F_{62}$ or $F_{25}F_{16}$, combined with $G_{34}G_{78}$. Thus the sequence of the different choices of $L_0$ for the six updating steps can equivalently be realized by relating the different steps by appropriate flavor rotations.

For the interaction part, i.e. the first term in eq.~\eqref{IN1}, we choose a different sequence of flavor rotations from one step to the next. For example, the transformation $-S_{xy}$ is now equivalently realized by the flavor rotation $F_{12}F_{65}G_{37}$. In short, we can realize the different choices of $L_0$ and $\Delta L$ for the six different steps by appropriate rotations in flavor space.

\subsection*{Symmetric continuum limit}
\medskip

At the present stage the naive continuum limit~\eqref{Z4} remains still rather asymmetric in flavor space. For example, the invariant $\chi^8$ is present only in the combinations with a time derivative. Also time and space appear in different roles. We can realize a higher degree of symmetry in space-time and in flavor space by continuing the six steps by another eighteen steps obtained by appropriate flavor rotations. The total sequence will then have $24$ steps. After these $24$ steps the sequence of updatings is repeated.

For $L_0$ step seven from $t+6\eps$ to $t+7\eps$ is obtained from step one by the flavor rotation $F_{17}F_{82}$ or $F_{18}F_{27}$. This exchanges the flavor pairs $(1,2)$ and $(7,8)$ with an additional minus sign, and therefore maps in eq.~\eqref{A10} $\Ttil_x^{12}+\Ttil_0^{78}\to-(\Ttil_0^{78}+\Ttil_x^{12})$. We combine this with the flavor reflection $G_{34}G_{56}$ which changes the sign of $\Ttil_y^{34}+\Ttil_z^{56}$. In the naive continuum limit~\eqref{Z4} this subtracts a term for which $\partial_t$ and $\partial_x$ are exchanged. For the interaction term the map from step one to step seven involves a different flavor rotation $F_{14}F_{85}$ or $F_{15}F_{48}$, combined with a reflection $G_{26}$ or $G_{37}$. Combined with the part from $L_0$ this subtracts from $\Delta L$ in eq.~\eqref{IN1} a term for which $\partial_t$ and $\partial_x$ are exchanged. Taken together, the continuum limit of $L=L_0+\Delta L$ is antisymmetric in the exchange $t\leftrightarrow x$. The following five steps obtain from step seven by the same sequence of reflections (or equivalent flavor rotations) $-S_{yz}$, $-S_{zx}$, $-S_{xy}$, $-S_{yz}$, $-S_{zx}$. Summing up the contributions of the first twelve steps subtracts from eq.~\eqref{Z4} a term for which $\partial_t$ and $\partial_x$ are exchanged. The expression $\sum_{n_s=0}^{11}L(t+n_s\eps)$ is antisymmetric in $\partial_t$ and $\partial_x$.

We repeat this procedure for the next six steps, performing now flavor rotations such that a further term is subtracted from $L$ for which $\partial_t$ and $\partial_y$ are exchanged. Finally, the last six steps involve flavor rotations such that a term with $\partial_t$ and $\partial_z$ exchanged is subtracted. In summary, the naive continuum limit for the sequence of twenty four steps reads
\bel{B1}
\Lbar(t)=\sum_{n_s=0}^{23}L(t+n_s\eps)=(8\eps)^4\eps^{\mu\nu\rho\sigma}A_{\mu\nu\rho\sigma}\ ,
\ee
with
\begin{align}
\label{B2}
A_{\mu\nu\rho\sigma}=\Big\{&\gl\chi^1\partial_\mu\chi^5-\chi^5\partial_\mu\chi^1\gr\gl\chi^2\partial_\nu\chi^6-\chi^6\partial_\nu\chi^2\gr\nn\\
\times&\gl\chi^3\partial_\rho\chi^7-\chi^7\partial_\rho\chi^3\gr\gl\chi^4\partial_\sigma\chi^8-\chi^8\partial_\sigma\chi^4\gr\nn\\
-&\gl\chi^1\partial_\mu\chi^2-\chi^2\partial_\mu\chi^1\gr\gl\chi^3\partial_\nu\chi^4-\chi^4\partial_\nu\chi^3\gr\nn\\
\times&\gl\chi^5\partial_\rho\chi^6-\chi^6\partial_\rho\chi^5\gr\gl\chi^7\partial_\sigma\chi^8-\chi^8\partial_\sigma\chi^7\gr\Big\}\ .
\end{align}
Here $\mu,\nu,\rho,\sigma=0,\dots3$, $\partial_0=\partial_t$, and $\eps^{\mu\nu\rho\sigma}$ is the four-index totally antisymmetric tensor, with $\eps^{0123}=-\eps^{1230}=1$. One recognizes the structure of the action~\eqref{I3}.

The final step of the continuum limit replaces for the action $S=\sum_{n_s=0}^{M_t}\sum_{x,y,z}L(t+n_s\eps,x,y,z)$ the sums by integrals, $\int_x=\int\text{d}t\,\text{d}x\,\text{d}y\,\text{d}z$,
\begin{align}
\label{B3}
S&=\sum_wL(w)=\frac1{8\cdot24}\int_x\eps^{-4}\Lbar(t,x,y,z)\nn\\
&=\frac{64}3\int_x\eps^{\mu\nu\rho\sigma}A_{\mu\nu\rho\sigma}\ .
\end{align}
This employs a lattice discretization of spacetime by the shifted lattice, $\int_x=8\eps^4\sum_w$, and a factor $1/24$ for the $24$ steps. The naive continuum limit yields indeed the diffeomorphism invariant action~\eqref{I3}. Eq.~\eqref{B2} fixes the flavor coefficient $G$ in eq.~\eqref{I3}. At this stage we have realized the construction of a probabilistic cellular automaton whose naive continuum limit coincides precisely with the action of spinor gravity in sect.~\ref{Sect. II}.

\subsection*{Higher derivatives}
\medskip

By construction, the naive continuum limit for the sequence of updatings realizes several discrete symmetries, as lattice rotations by $\pi/2$ and several reflections, completed by flavor rotations and reflections. These symmetries restrict the terms that can appear in the fermionic action in the naive continuum limit. In particular, it forbids terms with more than one derivative acting on an invariant $\chi^A$ that have been neglected so far. By virtue of these restrictions the naive continuum limit realizes symmetries beyond the discrete ones. This concerns continuous rotations in space and even diffeomorphism symmetry.

So far we have neglected possible terms involving two or more derivatives acting on a given invariant. They could arise from the omitted terms in eqs.~\eqref{C1},~\eqref{C2}. In principle, such terms contribute to the naive continuum limit. Each derivative comes with a factor $\eps$, such that for the naive continuum limit all terms with five or more derivatives vanish. We still need to discuss possible terms with up to four derivatives, for which at least two act on the same invariant. Since the interaction term $\Delta_L+L_0^4/24$ involves already four derivatives the possibly problematic terms can only arise from $L_0$. By virtue of discrete symmetries we will establish that no such terms appear in the naive continuum limit, thus confirming eq.~\eqref{B2}.

For the naive continuum limit of $L_0$ we can consider different flavor pairs separately. Consider first the flavors $1$ and $2$. 
The part $L_{0}^{(1,2)}$ is odd under $P_x$ and even under $P_{y}$, $P_{z}$, $T$. Furthermore, it is even under $S_{yz}$ and odd under the flavor exchange $1 \leftrightarrow 2$. It contains two factors of $\chi$ such that possible higher derivative terms beyond eq.~\eqref{C9} could be of the form $\partial_{y}\chi^{1}\partial_{y}\partial_{x}\chi^{2}-\partial_{y}\chi^{2}\partial_{y}\partial_{x}\chi^{1}$, and similar with $\partial_{y}$ replaced by $\partial_z$ or $§\partial_t$. (We identify terms related by partial integration.)
Another possibility is $\partial_{x}\chi^{1}\partial_{x}^{2}\chi^{2}-\partial_{x}\chi^{2}\partial_{x}^{2}\chi^{1}$. Further terms involve five or more derivatives and vanish in the naive continuum limit.

The sum of the first six terms in the sequence of updatings is by construction antisymmetric in $x\leftrightarrow y$, $x\leftrightarrow z$, $y\leftrightarrow z$. Adding the contributions of the different steps leaves a possible term
\begin{align}
L_{0,HD}^{(1,2)}\sim \; & (\partial_{z}^{2}-\partial_{y}^{2})\chi^{1}\partial_{x}\chi^{2} +
(\partial_{x}^{2}-\partial_{z}^{2})\chi^{1}\partial_{y}\chi^{2}\nn\\
&+ (\partial_{y}^{2}-\partial_{x}^{2})\chi^{1}\partial_{z}\chi^{2}-(1\leftrightarrow 2)\ ,
\end{align}
which is indeed odd under $S_{yz}$. On the other hand, $S_{yz}T_{x}=T_{x}$ implies that the higher derivative terms are even under $S_{yz}$ and we conclude $L_{0,HD}^{(1,2)}=0$. This holds similarly for  $L_{0,HD}^{(3,4)}$ and  $L_{0,HD}^{(5,6)}$.

For the flavor sector $(7,8)$ $L_0^{(7,8)}$ vanishes trivially since there are three terms with positive sign and three identical terms with negative sign. Finally, the other eighteen terms in the sequence have again a vanishing naive continuum limit of $L_0$ since they obtain from the first six terms by exchange of flavor pairs. In summary, for each individual term in the sequence of updatings higher derivative terms are expected for the naive continuum limit of $L_0$. It is the particular sequence of terms that implies a cancellation of the higher derivative terms in $L_0$ though the factor $\eps^{ijk}$ in eq.~\eqref{Z3}.

\subsection*{Signs of Grassmann basis elements and wave \\functions}
\medskip

The choice of the sequence of updatings involves particular minus signs which matter for the naive continuum limit. Replacing them by other signs would lead to a different naive continuum limit. One may wonder if this poses a problem for the meaning and the robustness of the naive continuum limit. We also admit arbitrary signs for the wave function of the cellular automaton, corresponding to arbitrary signs of the wave function for the fermionic quantum field theory. These signs do not matter for the probabilities $p_\tau=q_\tau^2$, but they play a role for other observables as the momentum observable which are represented by off-diagonal operators~\cite{CWPW}. In our formulation the choice of signs is related to a discrete local gauge symmetry which we briefly recapitulate here.

The choice of signs for the different updating steps is related to a particular choice of relative signs for the Grassmann basis elements which are used to implement the updating rule for the automaton in the fermionic formulation. In general, the choice of signs for the Grassmann basis elements is arbitrary. For the general bit-fermion map the association of bit configurations and Grassmann basis elements is only fixed up to a sign. A similar observation can be made for the relation between the real wave function and the probability distribution. The relation $p_{\tau}=q_{\tau}^{2}$ fixes $q_{\tau}$ only up to a sign. 

For our construction a change of sign of a Grassmann basis element can be associated to a change of sign of an element of the step evolution operator according to  eq.~\eqref{ST2}, or to a change of sign of a component of the wave function. The independence of physics from the choice of signs is reflected by a local symmetry of discrete $Z_{2}$-transformations.~\cite{CWPCA,CWNEW,FPCA}. Our particular choice of signs can be viewed as a particular choice of gauge. 

Making a gauge transformation to a different choice of signs is typically not compatible with a smooth continuum limit. Starting from a smooth wave function encoding the probabilistic information about the state of the automaton, and applying a local change of sign, typically leads to a discontinuity of the wave function or its derivatives. Our choice of signs leading to the naive continuum limit should be seen as a gauge choice which permits a description in terms of a smooth wave function. For the continuum limit it is sufficient that one choice of signs exists for which the wave function is sufficiently smooth.

The local $Z_2$-gauge transformations corresponding to sign-changes of the elements $q_\tau(t)$ act simultaneously on Grassmann basis elements, wave functions, step evolution operator and operators for observables. For a given choice of Grassmann basis functions and step evolution operator the gauge is fixed. Changing for this choice of gauge only the signs in the wave function leads to a different physical situation. It can be shown~\cite{CWPW} that arbitrary signs of the wave function lead to a positive overall probability distribution for all times.

\subsection*{ Coarse graining and continuum limit}
\medskip

Automata with a sequence of rotated or reflected updatings demonstrate important coarse graining properties of the continuum limit. For the naive continuum limit the order of the updatings plays no role. This corresponds to a coarse graining in time. The 24 time steps of a cycle are no longer resolved in their order - only an average over these time steps is considered. On the discrete level the order of the updating steps matters since the step evolution operators for the different steps do not commute. The association of the naive continuum limit with the true continuum limit therefore relies on the basic assumption that for sufficiently smooth wave functions this non-commutativity, which is formally of higher order in $\varepsilon$, does not play a role. This is similar to the situation for the Feynman path integral formulation of quantum mechanics. In the continuum limit the non-commutativity of the kinetic and potential parts in the Hamiltonian is neglected there. 

The naive continuum limit has a higher degree of symmetry as compared to the discrete formulation. In particular, it implements continuous rotation symmetry while only a discrete subgroup is respected by the lattice formulation. This situation is well known from lattice gauge theories or similar discrete formulations of continuum models. The question is if the renormalization flow of operators violating the enhanced symmetries drives them to zero in the infrared limit or not. 

With a given set of symmetries one expects only one or a few universality classes of the renormalization flow. The microscopic information beyond a few relevant parameters characterizing these universality classes is lost in the infrared limit. 
This induces the high predictive power of universal critical behavior. For the particular discretization of the continuous spinor gravity model that we have constructed in the present paper one may hope that it belongs indeed to the universality class with diffeomorphism symmetry. A dedicated renormalization group investigation of the discrete model will be required in order to settle this central question.

\section{Discussion\label{Sect. VI}}

We have constructed a probabilistic cellular automaton which is equivalent to a discrete quantum field theory for fermions. Its naive continuum limit is a model of spinor gravity in four dimensions. Quantum gravity with fermions is characterized by two important properties: the symmetries of diffeomorphisms (general coordinate transformations) and local Lorentz symmetry. While our automaton realizes invariance under local Lorentz transformations directly on the discrete level, the diffeomorphism symmetry shows up only in the naive continuum limit.

The updating rule for the automaton appears to be rather complicated if we combine 24 ``microscopic updatings" to a single step that is repeated. In a different view the microscopic steps are very simple, and the rule for a sequence of steps related by flavor rotations is also rather simple. It would not be difficult to implement this automaton numerically for sharp initial bit configurations. Realizing the evolution of a probability distribution over initial configurations will require more work due to the large number of possible initial configurations. We may consider our model as an example for an interesting discrete four-dimensional fermionic quantum field theory with interactions that can be mapped to a probabilistic cellular automaton. 

A key question concerns the true continuum limit of this model. The naive continuum limit is suggestive by neglecting terms O($\varepsilon$) in the continuum action. Qualitatively, this can be justified if the wave function changes only very little for time steps of the order of a few times $\varepsilon$ or for similar distances in space. In this respect it is not sufficient to start with a smooth initial wave function. The smoothness has to be preserved by the updating process. The conditions for this to happen are poorly known and we cannot assert at the present stage under which conditions the naive continuum limit constitutes a qualitatively correct guide.

For any continuum limit the probabilistic nature of the automaton or fermionic quantum field theory is crucial since only the continuity in the probabilistic information (wave function) permits to take $\varepsilon\to 0$. No continuum limit exists for the sharp wave function of a deterministic automaton. We have seen that taking the continuum limit corresponds to some type of coarse graining. On the coarse grained level the unique jump property of the automaton is lost. If individual time steps and their order can no longer be resolved, or if the information about the precise position in space of individual bits is no longer available, a given smoothed or averaged configuration of bits or fermions can develop into several distinct such configurations in a probabilistic way. We can still define an evolution operator on the coarse grained level. If no information is lost in the course of the time evolution on the coarse grained level, the evolution operator will be an orthogonal matrix in the real formulation, and unitary in the presence of a complex structure which is compatible with the evolution. Rows and columns do, however, contain now many non-zero elements. 

The feature of coarse graining is shared by the naive and the true continuum limit. The true continuum takes the effects of fluctuations into account. This leads to a dependence of effective couplings on the scale of the coarse graining (running couplings). It is therefore not guaranteed that the neglection of couplings which are proportional to powers of $\varepsilon$ on the microscopic level remains justified. The emergence of a continuum version of spinor gravity in the time continuum limit requires that terms violating rotation symmetry die out, similar to the continuum limit in lattice gauge theories. Furthermore, all terms violating diffeomorphism invariance of the coarse grained effective action have to die out as fluctuations on larger and larger scales are included consecutively. In the renormalization group language such symmetry violating couplings have to correspond to irrelevant parameters.

Even if our model could realize diffeomorphism invariance in the true continuum limit there remains still a long way to go before it can be accepted as a model of quantum gravity. It is necessary that composite vierbeins~\eqref{3A} and a composite metric, as $g_{\mu\nu}=\sum_{A}e_{\mu}{}^{Am}e_{\nu}{}^{An}\eta_{mn}$ or some other flavor structure, acquire a vacuum expectation value which defines a geometry. In the presence of such expectation values single fermions should propagate according to the Dirac equation in flat space, with a convenient generalization involving a composite spin connection in curved space. If this path toward gravity would reveal itself as successful, one could attempt to construct realistic grand unified models for particle physics coupled to quantum gravity. A generalization of the invariants $\chi^{A}$ to become also invariant under local SO(10)-gauge transformations with fermions in the fundamental spinor representation is straightforward. At present, we may only state that it seems not to be excluded that the Universe can be described by a probabilistic automaton.

\nocite{*} 
\bibliography{refs}
\end{document}